\documentstyle[12pt]{article}
\renewcommand{\theequation}{\arabic{equation}}
\newcommand{\EQ}{\begin{equation}}
\newcommand{\EN}{\end{equation}}

\newcommand{\ket}[1]{\left|#1\right\rangle}      

\newcommand{\bear}{\begin{eqnarray}}
\newcommand{\ear}{\end{eqnarray}}
\newcommand{\bt} { \begin{tabular} }
\newcommand{\et}{ \end{tabular} }
\newcommand{\bc} { \begin{center} }
\newcommand{\ec}{ \end{center} }

\newcommand{\btb} { \begin{table} }
\newcommand{\etb}{ \end{table} }

\begin{document}

\topmargin 0pt
\oddsidemargin 5mm
\newcommand{\NP}[1]{Nucl.\ Phys.\ {\bf #1}}
\newcommand{\PL}[1]{Phys.\ Lett.\ {\bf #1}}
\newcommand{\NC}[1]{Nuovo Cimento {\bf #1}}
\newcommand{\CMP}[1]{Comm.\ Math.\ Phys.\ {\bf #1}}
\newcommand{\PR}[1]{Phys.\ Rev.\ {\bf #1}}
\newcommand{\PRL}[1]{Phys.\ Rev.\ Lett.\ {\bf #1}}
\newcommand{\MPL}[1]{Mod.\ Phys.\ Lett.\ {\bf #1}}
\newcommand{\JETP}[1]{Sov.\ Phys.\ JETP {\bf #1}}
\newcommand{\TMP}[1]{Teor.\ Mat.\ Fiz.\ {\bf #1}}
     
\renewcommand{\thefootnote}{\fnsymbol{footnote}}
     
\newpage
\setcounter{page}{0}
\begin{titlepage}     
\begin{flushright}
\end{flushright}
\vspace{0.5cm}
\begin{center}
{\large The algebraic Bethe Ansatz for rational braid-monoid lattice models}\\
\vspace{1cm}
\vspace{1cm}
{\large $M.J.Martins^{1,2}$ and $P.B.Ramos^{2}$ } \\
\vspace{1cm}
{\em 1. Instituut voor Theoretische Fysica, Universiteit van Amsterdam\\
Valcknierstraat 65, 1018 XE Amsterdam, The Netherlands}\\
\vspace{.5cm}
{\em 2. Universidade Federal de S\~ao Carlos\\
Departamento de F\'isica \\
C.P. 676, 13560~~S\~ao Carlos, Brasil}\\
\end{center}
\vspace{0.5cm}
     
\begin{abstract}
In this paper we study isotropic integrable systems based on the braid-monoid 
algebra. These systems constitute a large family of rational multistate vertex
models and are realized in terms of the $B_n$, $C_n$ and $D_n$ Lie algebra
and by the superalgebra $Osp(n|2m)$. We present a unified formulation
of the quantum inverse scattering method for many of these lattice models.
The appropriate fundamental commutation rules are found, allowing us
to construct the eigenvectors and the eigenvalues of the transfer matrix
associated to the 
$B_{n}$, $C_{n}$, $D_{n}$, $Osp(2n-1|2)$, $Osp(2|2n-2)$, $Osp(2n-2|2)$ and 
$Osp(1|2n)$ models. The corresponding Bethe Ansatz equations can be formulated
in terms of the root structure of the underlying algebra.
\end{abstract}
\vspace{.15cm}
\centerline{PACS numbers: 05.20-y, 05.50+q, 04.20.Jb, 03.65.Fd}
\vspace{.1cm}
\centerline{Keywords: Algebraic Bethe Ansatz, Lattice Models}
\vspace{.15cm}
\centerline{January 1997}
\end{titlepage}

\renewcommand{\thefootnote}{\arabic{footnote}}
\section{Introduction}

	In this paper we look at the problem of diagonalization of the transfer 
matrix of a certain class of integrable two-dimensional lattice models. Their 
Boltzmann weights are intimately connected with a rational Baxterization of the 
braid-monoid algebra (see e.g. ref. \cite{WA} 
for a review) . In particular, we are going to analyse multistate 
vertex models which are based on the symmetries $B_{n}$, $C_{n}$, $D_{n}$, and 
$Osp(n|2m)$.

	One possible method of finding the eigenvalues of a given 
transfer matrix is by 
using the so-called analytical Bethe Ansatz \cite{RE}. This 
technique relies on the 
unitarity, crossing and analyticity properties of 
the transfer matrix and, in some cases, an extra amount of
phenomenological input is also required . This method has 
been applied to some of the 
models which we are going to consider in this 
paper, more precisely for the systems 
$B_{n}$, $C_{n}$, $D_{n}$ \cite{RE1,SU} and $Osp(1|2n)$ \cite{KU,MA}. 
Unfortunately, the explicit construction of eigenvectors 
of the transfer matrix is beyond the scope of 
the analytical Bethe Ansatz. The construction of exact eigenvectors, besides 
being an interesting problem on its own, is certainly an important 
step in the program 
of solving integrable systems. Thus, another 
route has to be taken if one 
wants to benefit from the knowledge of the eigenvectors. This would be the
case
of computing lattice correlation functions in the
framework developed by Izergin and Korepin \cite{KO1,KO}.

A more powerful mathematical method, based on first principles, is the 
quantum inverse scattering method \cite{FA}. This technique, together with the 
Yang-Baxter relation, offers us a unified viewpoint for studying the 
properties of integrability  of two-dimensional solvable models \cite{FA,KO,REV,KS}. One important 
feature of this method is that it permits us to present an algebraic formulation 
of the Bethe states. This step, however, depends much on our ability to 
disentangle the Yang-Baxter algebra in terms of appropriate commutation relations. The simplest structure of commutation rules has been discovered in 
the context of the $6-$vertex model \cite{FA} and its multi-state generalizations \cite{BA,KRA}.

	In this paper, we shall deal with the diagonalization problem of the 
transfer matrix of certain rational braid-monoid vertex models by means of 
the quantum inverse scattering method. We shall see that this program can be 
carried out in a universal way for a quite general class of 
systems: the $B_{n}$, $C_{n}$, $D_{n}$\footnote{We remark that the algebraic 
Bethe Ansatz solution for the $D_{n}$ vertex 
model has been previously discussed by de Vega and Karowiski 
in ref.\cite{KV} by using a different construction than the one pursued here (see 
further discussion in section 5) .}, $Osp(2n-1|2)$, 
$Osp(2|2n-2)$, $Osp(2n-2|2)$ 
and $Osp(1|2n)$ vertex models.  We show that the fundamental 
commutation rules have  a  common form 
in terms of the corresponding Boltzmann weights. As a 
consequence, the derivation of the eigenvectors, the eigenvalues and the 
associated Bethe Ansatz equations also have a quite general character for these 
vertex models. We believe that the unified picture proposed in this 
paper is a new result in the literature as well as the
Bethe Ansatz results for the superalgebra $Osp(n|2m)$. We remark that much
of our motivation, concerning such general picture, was prompted by our 
previous effort of presenting the algebraic Bethe Ansatz solution \cite{PM}
for the 4-dimensional representation of the supersymmetric $spl(2|1)$ vertex
model \cite{BRA,MAZ,BRA1,KUFA}.

	We have organized this papers as follows. In the 
next section, in order to 
make this paper self-contained, we review the basic properties of the braid-monoid
 algebra and its rational Baxterization. A convenient representation
 for Bethe Ansatz analysis is then presented for the $Osp(n|2m)$ symmetry. In 
section 3 we formulate the eigenvalue problem for the transfer matrix in terms of the 
quantum inverse scattering method. The fundamental commutation rules are explicitly 
exhibited. In section 4 we elaborate on the construction of the 
eigenvectors and eigenvalues of the $B_{n}$, $C_{n}$, $D_{n}$, $Osp(2n-1|2)$, 
 $Osp(2|2n-2)$, $Osp(2n-2|2)$ and $Osp(1|2n)$ vertex models and in section 5 the corresponding
 nested Bethe Ansatz equations are 
derived. These results allow us to conjecture the Bethe Ansatz equations of 
the general $Osp(n|2m)$ chain. Section 6 is reserved for our conclusion and 
remarks on the universal picture we have found for braid-monoid vertex 
systems. In appendices A and B we present details concerning the two and the 
three particle state, respectively. In appendix C we collect some useful 
relations for the supersymmetric formulation of the quantum inverse 
scattering method.

%
\section{The rational braid-monoid solution}

The braid-monoid algebra \cite{WA} is generated by the identity $I$, the
 braid operator $b_{i}$ and the monoid operator $E_{i}$ acting on the sites $i$ 
of a chain of length $L$. In general, these operators satisfy a set of
relations which goes by the name of Birman-Wenzel-Murakami 
algebra \cite{BWK,WA}. We recall that for
a braid operator $b_i$ 
we mean a object satisfying the braiding relation
( $b_i b_{i\pm1} b_i=b_{i\pm1}b_i b_{i\pm1}$ ) such that $b_i b_j = b_j b_i$
for $|i-j| \geq 2$. 
In this paper, we are interested in a degenerated
point of this algebra, when the braid operator $b_i$ and its inverse 
$b_i^{-1}$ are identical, i.e. $b_i = b_i^{-1} $. 
Here we choose the
braid operator as the graded 
permutation operator $b_{i} \equiv P_{i}^{g}$, defined by the 
following matrix elements \cite{KS}
\EQ
\left( P_{i}^{g} \right)_{ab}^{cd} = (-1)^{p(a)p(b)} \delta_{ad} \delta_{bc}
\EN
where $p(a)$ is the Grassmann parity of the $a$-th degree of freedom, 
assuming values $p(a)=0,1$.  
This is a generalization of the standard operation of permutation \cite{KS}, 
which 
distinguishes the `bosonic' $(p(a)=0)$ and the `fermionic' $(p(a)=1)$ degrees 
of freedom. The monoid $E_{i}$ is a Temperley-Lieb operator \cite{TL} 
and satisfies the relations
\EQ
\begin{array}{l}
E_{i}^{2} = {\cal K} E_{i} \\
E_{i}E_{i\pm1}E_{i} = E_{i} \\
E_{i}E_{j} = E_{j}E_{i} ; \hspace{2cm} |i-j| \geq 2 \\
\end{array}
\EN
where ${\cal K}$ is a $c$-number. The choice we 
have made for the braid operator ( see equation (1) )  
greatly simplifies the constraints 
between the braid and the monoid.  It is possible to show that the constraints
closing the braid-monoid algebra on such degenerated point are given by (
see e.g refs. \cite{MA1,XUE} )
\EQ
\begin{array}{l}
P_{i}^{g}E_{i} = E_{i}P_{i}^{g} = \hat{t} E_{i} \\
E_{i}P_{i\pm1}^{g}P_{i}^{g} = P_{i\pm1}^{g}P_{i}^{g}E_{i\pm1} = E_{i}E_{i\pm1} \\
\end{array}
\EN
where the constant $\hat{t}$ assumes only the values $\pm1$. Any other
constraint coming from the Birman-Wenzel-Murakami algebra \cite{WA,BWK}
can be derived from (3), from the braiding properties of
$P^g$, the fact that $(P_i^g)^2 =I $, and the Temperley-Lieb relations
(2). Lastly, we note
that these set of relations are invariant by the transformation 
$\hat{t} \rightarrow - \hat{t} $ and $ P_i^g \rightarrow - P_i^g $. 

	It turns out that the algebraic relations $(2)$ and $(3)$ can be 
`Baxterized' in terms of rational functions. In other words, it is possible 
to find a solution $R(\lambda)$ of the Yang-Baxter equation in terms of 
certain combination of the identity, $P_{i}^{g}$ and $E_{i}$. The solution 
comes in terms of rational functions \cite{MA1,XUE} and is given by\footnote{
In this paper the $R$-matrix is read as
$R(\lambda)= \displaystyle \sum_{abcd} R(\lambda)_{ab}^{cd} 
e_{ac} \otimes e_{bd}$, where the matrix elements of $e_{ab}$ are 
$[e_{ab}]_{ij}=\delta_{a,i} \delta_{b,j}$.}
\EQ
R(\lambda) = I + \lambda P^{g} - \frac{\lambda}{\lambda - \Delta} E
\EN
where $\Delta = \frac{(2-{\cal K}) \hat{t} }{2}$. This solution has a quasi-classical 
analog\footnote{The quasi-classical $r$-matrix can be obtained by  
redefining $\lambda \rightarrow \frac{\lambda}{\eta}$ and by expanding it around 
$\eta =0$.}, it is regular at $\lambda =0$ and also has a crossing point at 
$\lambda =  \Delta$. The next step is to search for explicit 
representations of the Temperley-Lieb operator $E_{i}$. In order to satisfy 
the braid-monoid restrictions $(2,3)$, one possible choice is to set the 
following Ansatz for the monoid $E_{i}$ \cite{MA1}
\EQ
(E_{i})_{ab}^{cd} = \alpha_{ab} \alpha^{-1}_{cd}
\EN
where $\alpha_{ab}$ are the elements of an invertible matrix. A quite general 
representation for matrix $\alpha$ can be found in the context of 
the $Osp(n|2m)$ symmetry. The integer $n$ and $2m$ stands for the number
of bosonic and fermionic degrees of freedom, respectively. The superalgebra
$Osp(n|2m)$ combines the orthogonal $O(n)$ and the sympletic $Sp(2m)$ 
Lie algebras ( see e.g. ref. \cite{CO} ), and its element $Z$ satisfies

\EQ
Z +M_{Osp} Z^{st} M_{Osp}^{-1} =0
\EN
where the symbol $st$ denotes the supertranspose 
operation  and the matrix $M_{Osp}$
is given by
\EQ
M_{Osp}=\left( \begin{array}{cc} 
	I_{n \times n} &   O_{n \times 2m} \\
	O_{2m \times n} & \left( \begin{array}{cc} O_{m \times m} 
& I_{m \times m} \\
	-I_{m \times m} & O_{m \times m} \\ \end{array} \right)\\
	\end{array}
	\right)
\EN
where $I_{k \times k}$ and $O_{k \times k}$ are the  
identity  and the null $k \times k$ matrices, respctively.  The 
elements of matrix (7)
can be used as an explicit representation \cite{MA1} for $\alpha_{ab}$,
i.e. $\alpha_{ab}= [M_{Osp}]_{ab} $, and the Temperly-Lieb parameter $\cal{K}$
is then fixed by
\EQ
{\cal K} = n-2m
\EN
For general values of $n$ and $m$, 
however, such representation breaks the $U(1)$ invariance of
the monoid $E_i$ \cite{MP} \footnote{ We remark that the $Osp(1|2m)$ is
an exception to this rule. In this case, a canonical transformation can
bring (7) in a $U(1)$ invariant form. }, and it is not appropriate for
Bethe Ansatz analysis. Usually, the lack of $U(1)$ invariance induces extra
difficulties on the formulation of a Bethe Ansatz, and therefore we
should look for other alternatives. This problem can be resolved
by the following construction. 
The monoid preserving the symmetry $U(1)$ is 
built in terms of an anti-diagonal matrix $\alpha$, whose elements 
are either  $+ 1$  or $-1$. 
The integer $m$ is the number of minus signs $(-1)$ and $n$ is the anti-trace 
of $\alpha$.  Furthermore, concerning the grading structure, the
elements $\alpha_{ab}$ only link degrees of freedom of
the same specie, namely $p(a)=p(b)$. The possible ways to 
distribute $\pm 1$ in the anti-diagonal, for a fixed ${\cal{K}}=n-2m$,  are 
then related by permutation of the grading indices (canonical transformations). 
However, one needs to make sure that $\hat{t}$, the elements $\alpha_{ab}$
and the parities $p(a)$ satisfy the braid-monoid relation (3), that is
$\hat{t} \alpha_{ab} = (-1)^{p(a)} \alpha_{ba} $. For example, one possible
representation for matrix $\alpha$ is the following block anti-diagonal
structure
\EQ
\alpha_{Osp(n|2m)}=\left( \begin{array}{ccc} 
	O_{m \times m} &   O_{m \times m} & {\cal{I}}_{n \times n} \\
	O_{m \times m} &   {\cal{I}}_{m \times m} & O_{n \times n} \\
	-{\cal{I}}_{m \times m} &   O_{m \times m} & O_{n \times n} \\
	\end{array}
	\right)
\EN
where ${\cal{I}}_{ k \times k} $ is a $ k \times k $ anti-diagonal matrix. 
In this case, the two compatible sequences of grading are $ f_1 \cdots
f_m b_1 \cdots b_n f_{m+1} \cdots f_{2m}$ for $\hat{t}=1$ and $b_1 \cdots
b_m f_1 \cdots f_n b_{m+1} \cdots b_{2m}$ for $\hat{t}=-1 $.  
One extra advantage of our construction 
is that  the vertex models $B_{n}$, $D_{n}$ and $C_{m}$ can be nicely 
represented in terms of the limits $m \rightarrow 0$ and
$ n \rightarrow 0$, respectively. More precisely, we have
\EQ
\alpha_{B_n} = {\cal{I}}_{2n+1 \times 2n+1 },~~ \alpha_{D_n} =
{\cal{I}}_{ 2n \times 2n },~~  
\alpha_{C_m}=\left( \begin{array}{cc} 

	O_{m \times m} &     {\cal{I}}_{m \times m} \\
	-{\cal{I}}_{m \times m} &   O_{m \times m}  \\
	\end{array}
	\right)
\EN

In this paper, besides the  $B_{n}$, $C_{n}$ and $D_{n}$ models, we are 
primarily interested in the supersymmetric models related to the 
$Osp(2n-1|2)$, $Osp(2|2n-2)$,$Osp(2n-2|2)$ and 
$Osp(1|2n)$ symmetries. The first reason is 
because, as we shall see below, the nested Bethe Ansatz 
formulation for the $Osp(2n-1|2), Osp(2|2n-2)$ and
$Osp(2n-2|2)$ models goes fairly parallel to that of the
$B_{n}$, $C_{n}$ and $D_{n}$ vertex models, respectively. 
Secondly, because they  
exhaust the main basic symmetries present 
in the  general $Osp(n|2m)$ superalgebra. In order to see
these relations, we show in figure 1 the Dynkin diagrams of the
superalgebra $Osp(n|2m)$ as well as those of the Lie algebras $B_n$, $C_n$ and
$D_n$. We notice that the $Osp(1|2n)$ superalgebra has a special 
structure of roots. In fact, such special character will be
present in many points of our Bethe Ansatz analysis of the $Osp(1|2n)$ vertex
models. 

We turn now to the analysis of the Boltzmann weights of these 
vertex models. In general, these are multistate vertex models having 
one of $q$  possible states 
on each bond of the two dimensional square lattice.
The functional form of the Boltzmann weights depends
directly on the values of $\hat{t}$, $\cal{K}$. For the $Osp(n|2m)$ vertex
models, the weights also depend on the sequence of grading that has been chosen. In table 1
we have collected the values of $q$, $\hat{t}$ and $\cal{K}$ for the
vertex models 
$B_{n}$, $C_{n}$, $D_{n}$, $Osp(2n-1|2)$,
$Osp(2|2n-2)$, $Osp(2n-2|2)$ and $Osp(1|2n)$. Moreover, for the
first three supersymmetric models we have used the grading $fb \cdots bf$ and
for $Osp(1|2n)$ we have taken the sequence $b \cdots bfb \cdots b$. The basic
reason for this choice is because it uses the minimum number of fermionic
degrees of freedom in the grading sequence compatible with number 
$\cal{K}$ and our choice of
$\hat{t}$ (see table 1), and therefore saving the commutation rules of
the presence of many extra minus signs.
We remark that the structure of equation (9) is compatible with the
above mentioned grading sequences for the $Osp(2n-1|2)$, $Osp(2n-2|2)$ and
$Osp(1|2n)$ vertex models. For the $Osp(2|2n-2)$ system, however, we still
have to perform an extra transformation in matrix (9),
by exchanging the least $-1$ and the $(n+1)$-th $+1$ degrees of
freedom. More
specifically, we have that $\alpha_{Osp(2|2n-2)}= anti-diagonal( 
1_1, \cdots 1_{n},-1_{n+1},-1_{n+2}, \cdots,-1_{2n-1},
1_{2n} ) $. This transformation is quite helpful because 
it allows us to treat the nesting
problem for the $C_n$ and $Osp(2|2n-2)$ models 
in a common way. 

Taking into account these considerations, we find
that there are only  few possible distinct functional forms for the
Boltzmann weights. 
In fact, as we shall see in next section, our algebraic formulation will 
require at most five distinct functional dependences. We name these Boltzmann 
weights by $a(\lambda)$, $b(\lambda)$, $c_{n}(\lambda)$, $d_{n}(\lambda)$, $e_{n}(\lambda)$ and 
they are summarized on Table 2. To some extent, this is the miraculous fact 
of integrability, which in our approach is encoded in the commutation 
rules to be presented in next section.

\section{The eigenvalue problem and the commutation rules}

We start this section by describing the eigenvalue problem for the 
transfer matrix $T(\lambda)$ associated to the vertex models defined in section 
2. We are interested to determine the eigenvalues and the eigenvectors of 
$T(\lambda)$ on a square lattice of size $L \times L$. The diagonalization problem is defined by
\EQ
T(\lambda) \ket{\Phi} = \Lambda (\lambda) \ket{\Phi}
\EN

	An important object in the quantum inverse scattering method 
\cite{FA,KO,REV,KS} is the Yang-Baxter algebra of the monodromy matrices. The 
monodromy matrix ${\cal T}(\lambda)$ acts on the tensor product of an 
auxiliary space ${\cal A} \equiv {\cal C}^{q}$ and on the quantum Hilbert 
space $ {\cal C}^{Lq}$. The transfer matrix ${\cal T}(\lambda)$ is the trace of
the monodromy matrix ${\cal T}(\lambda)$ over the auxiliary space 
${\cal A}$, i.e., $T(\lambda) = Tr_{{\cal A}} {\cal T}(\lambda)$. A convenient 
way of writing the monodromy matrix is in terms of a product of 
operators ${\cal L}_{{\cal A}i}(\lambda)$ as 
\EQ
{\cal T}(\lambda) = {\cal L}_{{\cal A}L}(\lambda){\cal L}_{{\cal A}L-1}(\lambda)....{\cal L}_{{\cal A}1}(\lambda)
\EN
where ${\cal L}_{{\cal A}i}(\lambda)$ are $q \times q$ matrices acting
on the lattice sites $i=1, \cdots, L$ 
whose elements are operators on the Hilbert space ${\cal C}^{Lq}$. 
The elements of the vertex operator ${\cal L}_{ab}^{cd}(\lambda)$ are 
related to those of the $R$-matrix (4) by a permutation on the ${\cal C}^{q} 
\times {\cal C}^{q}$ tensor space as 
\EQ
{\cal L}_{ab}^{cd}(\lambda) =  R_{ba}^{cd}(\lambda)
\EN
and the Yang-Baxter algebra for two monodromy matrices with distinct spectral 
parameters reads
\EQ
R(\lambda - \mu) {\cal T}(\lambda) \otimes {\cal T}(\mu) =
{\cal T}(\mu) \otimes {\cal T}(\lambda) R(\lambda - \mu)
\EN

	The intertwining equation (14) is the corner-stone of the quantum 
inverse scattering approach, allowing us to derive the fundamental commutation 
relations. Unfortunately, there is no recipe to immediately find the 
appropriate commutation rules from the Yang-Baxter algebra. Much of the 
insight comes from the properties of the vertex operator ${\cal L}_{{\cal A}i}(\lambda)$ 
itself and from the reference state we choose to begin the construction of the 
Hilbert space. As a reference state we take the standard ferromagnetic 
pseudovacuum given by 	
\EQ
\ket{0} = \prod_{i=1}^{L} \otimes \ket{0}_{i} , ~~
\ket{0}_{i} = 
\pmatrix{
1 \cr
0 \cr
\vdots \cr
0 \cr}_{q}
\EN
where $q$ stands for the length of the vectors $\ket{0}_{i}$. The vertex 
operator ${\cal L}_{{\cal A}i}(\lambda)$, when acting on state  $\ket{0}_{i}$, 
has the following triangular property, i.e.,
\EQ 
{\cal L}_{{\cal A}i}(\lambda)\ket{0}_{i} = 
\pmatrix{
a(\lambda) \ket{0}_i &  *  &  *  &  * & \dots & * & *  \cr
0  &  b(\lambda) \ket{0}_i  &  0  &  0 & \dots & 0 & *  \cr
0  &  0  &  b(\lambda) \ket{0}_i  &  0 & \dots & 0 & *  \cr
\vdots & \vdots & \vdots & \ddots & \dots & \vdots & \vdots \cr
0  &  0  &  0  & 0 & \dots & b(\lambda)\ket{0}_i & * \cr
0  &  0  &  0  & 0 & \dots & 0 &  e_{n}(\lambda) \ket{0}_i  \cr}_{q \times q}
\EN
where the symbol $*$ represents some values that are not necessary to 
evaluate explicitly for further discussion. The next step is to write an appropriate 
Ansatz for the matrix representation of ${\cal T}(\lambda)$ on the 
auxiliary space ${\cal A}$. The triangular property $(16)$ suggests us to seek 
for the following structure
\EQ
{\cal T}(\lambda) = 
\pmatrix{
B(\lambda)       &   \vec{B}(\lambda)   &   F(\lambda)   \cr
\vec{C}(\lambda)  &  \hat{A}(\lambda)   &  \vec{B^{*}}(\lambda)   \cr
C(\lambda)  & \vec{C^{*}}(\lambda)  &  D(\lambda)  \cr}_{q \times q}
\EN
where $\vec{B}(\lambda)$,  $\vec{B^{*}}(\lambda)$,
$\vec{C}(\lambda)$, $\vec{C^{*}}(\lambda)$  are two component vectors 
with dimensions $1 \times (q-2)$, $(q-2) \times 1$, $(q-2) \times 1$, $1 \times (q-2)$, respectively. The 
operator $\hat{A}(\lambda)$ is a $(q-2) \times (q-2)$ 
matrix and we denote its elements by $A_{ab}(\lambda)$. The other remaining 
operators $B(\lambda)$, $C(\lambda)$, $F(\lambda)$, 
and $D(\lambda)$ are scalars. Putting them all together, we then have
a $(q \times q)$ matrix representation for the 
monodromy matrix ${\cal T}(\lambda)$. Taking into account these definitions, 
the diagonalization problem (11) for the transfer matrix $T(\lambda)$ becomes 
\EQ
[B(\lambda)+\sum_{a=1}^{q-2}A_{aa}(\lambda)+D(\lambda)] \ket{\Phi} = 
\Lambda(\lambda) \ket{\Phi}
\EN
	As a consequence of definitions (12,17) and 
the triangular property (16), 
we find that the diagonal operators of ${\cal T}(\lambda)$ satisfy the 
following relations
\EQ
B(\lambda)\ket{0} = [a(\lambda)]^{L}\ket{0};~~ D(\lambda)\ket{0} = 
[e_{n}(\lambda)]^{L}\ket{0};~~ A_{aa}(\lambda)\ket{0} = [b(\lambda)]^{L}\ket{0} , a=1, \dots , q-2
\EN
as well as the annihilation properties
\EQ
C_{a}(\lambda)\ket{0} = 0;~~ C_{a}^{*}(\lambda)\ket{0} = 0;~~ 
C(\lambda) \ket{0} = 0; ~~ A_{ab}(\lambda)\ket{0} = 0,~(a,b =1, \cdots , q-2; a \neq b)
\nonumber \\
\EN

	This means that the reference state is an exact and trivial eigenvector 
with eigenvalue
\EQ
\Lambda(\lambda) = [a(\lambda)]^{L} + (q-2) [b(\lambda)]^{L} + 
[e_{n}(\lambda)]^{L}
\EN
and also that the fields $\vec{B}(\lambda)$,  $\vec{B^{*}}(\lambda)$ and 
 $F(\lambda)$ should play the role of the creation operators on the 
reference state. In order to construct the full Hilbert space we  need to
find the commutation relations between the creation, diagonal and 
annihilation fields. In principle, all the information concerning commutation 
rules are encoded in the integrability condition (14). The basic 
problem is to collect them in a 
convenient form. For instance, for the $6$-vertex 
model \cite{FA,KO} and its multi-state generalizations \cite{REV,BA,KRA} they 
come almost directly, after substituting the appropriate form for ${\cal T}(\lambda)$
 and the associated $R$-matrix on the Yang-Baxter algebra (14). For the 
models we intend to analyse in this paper, that is 
the rational braid-monoid vertex 
models (4), this is not the case and some additional work is necessary. For example,
 in order to get the `nice' commutation rule between the creation 
operator $\vec{B}(\lambda)$ and the diagonal field $\hat{A}(\lambda)$ we also 
have to disentangle the commutation relations between $B(\mu)$ and 
$\vec{B^{*}}(\lambda)$. The trick goes much along the lines we have already 
explained in details for the $spl(2|1)$ supersymmetric vertex model
\cite{PM}. 
It turns out that these ideas can be generalized for the
$B_{n}$, $C_{n}$, $D_{n}$, $Osp(2n-1|2)$, 
$Osp(2|2n-2)$, $Osp(2n-2|2)$ and $Osp(1|2n)$ 
multi-state vertex models, and therefore 
we have been able to derive their
fundamental commutation rules.
As stressed below, the main procedure is quite cumbersome, and 
here we just list our final results for the most important commutation rules. 
Between the diagonal operators $\hat{A}(\lambda)$, $B(\lambda)$, $D(\lambda)$ 
and the creation field $\vec{B}(\lambda)$ we have
\bear
\hat{A}(\lambda) \otimes \vec{B}(\mu) = 
\frac{1}{b(\lambda - \mu)}[\vec{B}(\mu) \otimes \hat{A}(\lambda) ]. 
{\tilde{X}}^{(1)}(\lambda-\mu)
- \frac{1}{b(\lambda-\mu)} \vec{B}(\lambda) \otimes \hat{A}(\mu)+   
\nonumber \\
\frac{d_{n}(\lambda-\mu)}{e_{n}(\lambda-\mu)} \left[
 \vec{B^{*}}(\lambda)B(\mu) 
+\frac{1}{b(\lambda-\mu)}F(\lambda)\vec{C}(\mu)  
-\frac{\{ 1-\hat{t} b(\lambda-\mu) \} }{b(\lambda-\mu)}F(\mu)\vec{C}(\lambda) \right ] 
\otimes \vec{\xi}
\ear
\EQ
B(\lambda)\vec{B}(\mu) = 
\frac{a(\mu-\lambda)}{b(\mu-\lambda)} \vec{B}(\mu)B(\lambda) - 
\frac{1}{b(\mu-\lambda)} \vec{B}(\lambda)B(\mu),
\EN
\bear
D(\lambda)\vec{B}(\mu) = 
\frac{b(\lambda-\mu)}{e_{n}(\lambda-\mu)} \vec{B}(\mu)D(\lambda) 
+ \frac{1}{e_{n}(\lambda-\mu)} F(\mu)\vec{C^{*}}(\lambda)
 \nonumber \\ 
 - \frac{c_{n}(\lambda-\mu)}{e_{n}(\lambda-\mu)} F(\lambda)\vec{C^{*}}(\mu)
- \frac{d_{n}(\lambda-\mu)}{e_{n}(\lambda-\mu)} 
\vec{\xi} . \{ \vec{B^{*}}(\lambda) \otimes \hat{A}(\mu) \}.
\ear
where ${\tilde{X}}^{(1)}(\lambda)$  is a
factorizable auxiliary $R$-matrix responsible for the $first$
nested Bethe Ansatz structure. In what follows we shall
see that it is also useful to introduce a second factorizable $R$-matrix 
$X^{(1)}(\lambda)$. The matrix elements of $X^{(1)}$ and 
${\tilde{X}}^{(1)}(\lambda)$, however,  are 
related to each other under permutation
of their horizontal and vertical spaces, namely $X^{(1)}(\lambda)^{ab}_{cd} =
{\tilde{X}}^{(1)}(\lambda)^{cd}_{ab} $. This distinction is
necessary in order to include the 
$Osp(1|2n)$ model in our discussion. This model
is an exception because its auxiliary $R$-matrix ${\tilde{X}}^{(1)}(\lambda)$
is no 
longer $T$-invariant.  All the other vertex models, however, have the 
property 
$X^{(1)}(\lambda)^{ab}_{cd} =
X^{(1)}(\lambda)^{cd}_{ab} $, and therefore $X^{(1)}(\lambda)$ and
${\tilde{X}}^{(1)}(\lambda)$ are indeed identical. 
In table 3 we have collected the Boltzmann weights of the $R$-matrix $X^{(k)}(
\lambda)$ on a certain $k$-th step of the nested Bethe Ansatz. We also
show the value of crossing parameter $\Delta^{(k)}$ corresponding to $X^{(k)}(\lambda)$. 
From this table we notice 
that the pairs of models:  $ \{ B_{n}$,$ Osp(2n-1|2) \}$, 
$\{ C_{n}$, $Osp(2|2n-2) \} $,
and $\{ D_{n}$, $Osp(2n-2|2) \}$ share the same 
auxiliary $X^{(k)}(\lambda)$ matrix. In fact, this is not 
that surprising if one
takes into account the similarities between their root 
structure ( see figure 1).
Furthermore, for these models, the matrix
$X^{(k)}$ is precisely the $R$-matrix associated to the $B_{n-k}$, $C_{n-k}$
and $D_{n-k}$ vertex models, respectively. However, since 
the vertex models on a given pair have distinct Boltzmann weights,  we shall
see that their
eigenvectors and eigenvalues will be in fact different too. As before, the 
$Osp(1|2n)$ model is an exception. Here, the nesting still keeps the same
structure as the original model and we have the 
embedding $Osp(1|2n) \subset Osp(1|2n-2) \subset \cdots 
\subset Osp(1|2)$.  Finally, the
vector $\vec{\xi}$ and its ``conjugated'' $\vec{{\xi}^{*}}$
are given in terms of the matrix 
$\alpha$, defining  the monoid operator present 
in the $R$-matrix $X^{(1)}$. They are defined by
\EQ
\vec{\xi} = \sum_{a,b=1}^{q-2} \alpha^{-1}_{ab}  
(\hat{e}_{a} \otimes \hat{e}_{b}), \;
\vec{{\xi}^{*}} = \sum_{a,b=1}^{q-2} \alpha_{ab}  
(\hat{e}_{a} \otimes \hat{e}_{b})
\EN
where $\hat{e}_{i}$ denotes the 
elementary projection on the $i$-th position. 

We now will give other important commutation relations. 
The commutation rules between the scalar creation operator $F(\lambda)$ and the 
diagonal fields are given by
\bear
\hat{A}(\lambda)F(\mu) = 
[1 - \frac{1}{b^{2}(\lambda-\mu)}] F(\mu)\hat{A}(\lambda)
+ \frac{1}{b^{2}(\lambda-\mu)} F(\lambda) \hat{A}(\mu)
\nonumber \\
 - \frac{1}{b(\lambda-\mu)} \left[
 \vec{B}(\lambda) \otimes \vec{B^{*}}(\mu) 
- \vec{B^{*}}(\lambda) \otimes \vec{B}(\mu) \right]
\ear
\EQ
B(\lambda)F(\mu) = 
\frac{a(\mu-\lambda)}{e_{n}(\mu-\lambda)} F(\mu)B(\lambda) - 
\frac{c_{n}(\mu-\lambda)}{e_{n}(\mu-\lambda)} F(\lambda)B(\mu)
- \frac{d_{n}(\mu-\lambda)}{e_{n}(\mu-\lambda)} 
\vec{{\xi}^{*}} .  \{ \vec{B}(\lambda) \otimes \vec{B}(\mu) \}
\EN
\EQ
D(\lambda)F(\mu) = 
\frac{a(\lambda-\mu)}{e_{n}(\lambda-\mu)} F(\mu)D(\lambda) - 
\frac{c_{n}(\lambda-\mu)}{e_{n}(\lambda-\mu)} F(\lambda)D(\mu)
- \frac{d_{n}(\lambda-\mu)}{e_{n}(\lambda-\mu)} \vec{\xi} . 
\{ \vec{B^{*}}(\lambda) \otimes \vec{B^{*}}(\mu) \}
\EN
and the commutation relations between the creation operators are 
\bear
\vec{B}(\lambda) \otimes \vec{B}(\mu) = \frac{1}{a(\lambda-\mu)}
[ \vec{B}(\mu) \otimes \vec{B}(\lambda) ]. X^{(1)}(\lambda-\mu)
\nonumber \\
+\frac{d_{n}(\lambda-\mu)}{e_{n}(\lambda-\mu)}  
\left[ F(\lambda)B(\mu) - \frac{\{ 1-\hat{t} b(\lambda-\mu) \}
 }{a(\lambda-\mu)}F(\mu)B(\lambda) \right] \vec{\xi}
\ear
\EQ
\left[ F(\lambda), F(\mu) \right] = 0
\EN
\EQ
F(\lambda) \vec{B}(\mu) = \frac{1}{a(\lambda -\mu)} F(\mu) \vec{B}(\lambda) + 
 \frac{b(\lambda-\mu)}{a(\lambda -\mu)} \vec{B}(\mu) F(\lambda)
\EN
\EQ
\vec{B}(\lambda) F(\mu) = \frac{1}{a(\lambda -\mu)} \vec{B}(\mu) F(\lambda) + 
 \frac{b(\lambda-\mu)}{a(\lambda -\mu)} F(\mu) \vec{B}(\lambda)
\EN

We close this section with the following remark. In this section, we have
kept our presentation concerning the basic properties of the quantum inverse
scattering approach as general as possible. All the vertex models have been
treated by the same and standard way of formulating the quantum 
inverse scattering method. We notice, however, that the solution of the
supersymmetric $Osp(n|2m)$ models could also be presented in  terms of a 
graded framework from the very beginning. This is possible because the $R$-matrix (4) has a null Grassmann parity, and consequently can produce a vertex 
operator
${\cal{L}}_{{\cal{A}}i}$ satisfying either 
the Yang-Baxter equation (standard formulation ) 
or its graded version \cite{KS,KU}.
The last choice, however, is the most natural way of formulating 
the vertex
operator 
${\cal{L}}_{{\cal{A}}i}$  
for the $Osp(n|2m)$ models, if one wants to make a real distinction between
the bosonic and fermionic degrees of freedom. In other words, the graded
quantum inverse method makes sure that the fermionic degrees of 
freedom anticommutes even if they act on different lattice sites.
In order to accomplish this ``non-local'' property, one has 
to use the supersymmetric (graded) formalism
developed in refs. \cite{KU,KS} ( see also refs. \cite{KE} ), which basically
changes the trace and the tensor product properties by their analogs on the
graded space. In appendix $C$ we have summarized this approach for the
$Osp(2n-1|2)$, $Osp(2|2n-2)$, $Osp(2n-2|2)$ and $Osp(1|2n)$ vertex models.
We notice, however, that the supersymmetric formulation does not simplify
the original problem of the diagonalization of the corresponding 
transfer matrices. We shall come back to this point again in section 5, where
the final results for the eigenvalues and Bethe Ansatz equations are presented.
\section{The eigenvectors and the eigenvalue construction}

In the quantum inverse scattering scheme the eigenvectors are 
constructed by acting the creation operators on the ferromagnetic pseudovacuum 
 $\ket{0}$. The excitations over the pseudovacuum $\ket{0}$ present a 
multi-particle feature, characterized by a set of variables 
$\{\lambda^{(1)}_{1}, \dots , \lambda^{(1)}_{m_{1}}\}$ which are determined after a 
posteriori analysis. The way we are going to start our discussion is much 
inspired from our previous 
results for the supersymmetric $spl(2|1)$ vertex model 
\cite{PM}. Indeed, the construction we shall present here is a  non-trivial 
generalization of ideas  discussed by us in ref. \cite{PM}. Due to the presence 
of many kinds of 
creation fields, it is convenient to represent the $m_1$-particle 
state $\ket{\Phi_{m_1}(\lambda^{(1)}_{1}, 
\cdots , \lambda^{(1)}_{m_{1}})}$ by the 
following linear combination
\EQ
\ket{\Phi_{m_{1}}(\lambda^{(1)}_{1}, \cdots , \lambda^{(1)}_{m_{1}})} =  
\vec {\Phi}_{m_{1}}(\lambda^{(1)}_{1}, \cdots , \lambda^{(1)}_{m_{1}}).\vec{{\cal F}} \ket{0}
\EN
where $\vec {\Phi}_{m_{1}}(\lambda^{(1)}_{1}, 
\cdots , \lambda^{(1)}_{m_{1}})$ and 
$\vec{{\cal F}}$ are multi-component 
vectors with $(q-2)^{m_{1}}$ components. The 
dependence on the creation fields is encoded in the vector 
 $\vec {\Phi}_{m_{1}}(\lambda^{(1)}_{1}, \cdots , \lambda^{(1)}_{m_{1}})$ and 
$\vec{{\cal F}}$ is a constant in this space of fields. We shall denote the 
components of $\vec{{\cal F}}$ by ${\cal F}^{a_{m_{1}} \dots a_{1}}$.

The simplest excitation, i.e., the one-particle state, can be built 
only in terms of the creation fields $\vec{B}(\lambda)$, namely
\EQ
\vec {\Phi}_{1}(\lambda^{(1)}_{1}) = \vec{B}(\lambda^{(1)}_{1})
\EN
and as a consequence of equations (33,34), the one-particle state is given by
\EQ
\ket {\Phi_{1}(\lambda^{(1)}_{1})} = B_{a}(\lambda^{(1)}_{1}) {\cal F}^{a} \ket{0} 
\EN
where here and in the following repeated indices denote the sum  over 
$a=1, \cdots ,(q-2)$. The commutation relations $(22-24)$ and properties 
$(19,20)$ can be used to solve the eigenvalue problem for 
$\ket{\Phi_{1}(\lambda^{(1)}_{1})}$. The solution is based 
on the following relations
\EQ
B(\lambda) \ket{\Phi_{1}(\lambda^{(1)}_{1})} = 
\frac{a(\lambda^{(1)}_{1}-\lambda)}{b(\lambda^{(1)}_{1}-\lambda)} 
[a(\lambda)]^L \ket{\Phi_{1}(\lambda^{(1)}_{1})}
 - \frac{1}{b(\lambda^{(1)}_{1}-\lambda)} [a(\lambda^{(1)}_{1})]^L B_{a}(\lambda)F^{a} \ket{0}
\EN
\EQ
D(\lambda) \ket{\Phi_{1}(\lambda^{(1)}_{1})} =
\frac{b(\lambda-\lambda^{(1)}_{1})}{e_{n}(\lambda-\lambda^{(1)}_{1})} [e_{n}(\lambda)]^L 
\ket{\Phi_{1}(\lambda^{(1)}_{1})}
- \frac{d_{n}(\lambda-\lambda^{(1)}_{1})}
{e_{n}(\lambda-\lambda^{(1)}_{1})} [b (\lambda^{(1)}_{1})]^L \xi_{ab}B^{*}_{a}(\lambda)F^{b} \ket{0}
\EN
\bear
&& \sum_{a=1}^{q-2}  A_{aa}(\lambda) \ket{\Phi_{1}(\lambda^{(1)}_{1})} = 
\frac{1}{b(\lambda-\lambda^{(1)}_{1})} \left[ 1+(q-2) (\lambda-\lambda^{(1)}_{1}) - \frac{(\lambda-\lambda^{(1)}_{1})}{(\lambda-\lambda^{(1)}_{1}+ \Delta^{(1)})} \right]
[b(\lambda)]^L \ket{\Phi_{1}(\lambda^{(1)}_{1})}
\nonumber \\
&&  - \frac{1}{b(\lambda-\lambda^{(1)}_{1})} [b(\lambda^{(1)}_{1})]^L B_{a}(\lambda)F^{a} \ket{0} 
+ \frac{d_{n}(\lambda-\lambda^{(1)}_{1})}
{e_{n}(\lambda-\lambda^{(1)}_{1})} [a(\lambda^{(1)}_{1})]^L \xi_{ab} B^{*}_{a}(\lambda)F^{b} \ket{0} 
\ear

The terms proportional to the eigenstate $\ket{\Phi_{1}(\lambda^{(1)}_{1})}$ 
are denominated `wanted' terms and 
contribute to the eigenvalue  $\Lambda(\lambda,\lambda^{(1)}_{1})$. The 
remaining ones are the so called `unwanted' terms and they have to be canceled 
by imposing further restriction on variable $\lambda^{(1)}_{1}$. Such 
constraint goes by the name of Bethe Ansatz equation. From relations $(36-38)$,
 for the one-particle state, we find a single equation given by\footnote{ This 
means that the degeneracy of the one-particle state is $(q-2)$, because no 
further constraint is necessary for vector $\vec{{\cal F}}$.}
\EQ
\left [\frac{a(\lambda^{(1)}_{1})}{b(\lambda^{(1)}_{1})}\right ]^{L} = 1
\EN

The feature of the two-particle state is a bit more complicated. Now, the 
scalar operator $F(\lambda)$ starts to play an important role on the 
eigenvector construction. This becomes clear if one takes into account 
the commutation rule between two creation fields of type $\vec{B}(\lambda)$ 
(see equation (29)). For the two-particle state we have to seek for a 
combination between two fields of type ${\vec{B}(\lambda)}$ with the scalar field 
$F(\lambda)$. The structure of commutation rule $(29)$ suggests us 
to take  the 
following combination for the vector $\vec {\Phi}_{2}(\lambda^{(1)}_{1},\lambda^{(1)}_{2})$
\EQ
\vec {\Phi}_{2}(\lambda^{(1)}_{1},\lambda^{(1)}_{2}) = 
\vec{B}(\lambda^{(1)}_{1}) 
\otimes \vec{B}(\lambda^{(1)}_{2}) + 
\hat{h}(\lambda^{(1)}_{1},\lambda^{(1)}_{2}) 
F(\lambda^{(1)}_{1}) B(\lambda^{1}_{2}) \vec{\xi}
\EN

One way of fixing the 
function $\hat{h}(\lambda^{(1)}_{1},\lambda^{(1)}_{2})$ is to
 notice that the eigenvalue problem for the two-particle state generates 
certain kind of `unwanted' terms which are proportional to $\vec{\xi}.\vec{{\cal F}}$. 
They are given by
\EQ
F(\lambda) D(\lambda^{(1)}_{1}) B(\lambda^{(1)}_{2}) ; ~~
\vec{B}(\lambda).\vec{B^{*}}(\lambda^{(1)}_{1}) B(\lambda^{(1)}_{2}) ;~~
\vec{\xi}.[\vec{B^{*}}(\lambda) \otimes \vec{B^{*}}(\lambda^{(1)}_{1})] B(\lambda^{(1)}_{2}) 
\EN

We call these structures `easy unwanted' terms, because they can be 
automatically canceled out by an appropriate choice of the form of the function 
$\hat{h}(\lambda^{(1)}_{1},\lambda^{(1)}_{2})$, namely
\EQ
\hat{h}(\lambda^{(1)}_{1},\lambda^{(1)}_{2}) = 
\hat{h}(\lambda^{(1)}_{1}-\lambda^{(1)}_{2}) =
- \frac{d_{n}(\lambda^{(1)}_{1}-\lambda^{(1)}_{2})}{e_{n}(\lambda^{(1)}_{1}-\lambda^{(1)}_{2})}
\EN

There is a more elegant way, however, to determine the function 
$\hat{h}(\lambda^{(1)}_{1},\lambda^{(1)}_{2})$. In general, we expect that the 
extra constraints on variables ${\lambda^{(1)}_{i}}$, i.e. the Bethe Ansatz 
equations, are invariant under index permutation. This means, for the 
two-particle, that vector $\vec {\Phi}_{2}(\lambda^{(1)}_{1},\lambda^{(1)}_{2})$
 and $\vec {\Phi}_{2}(\lambda^{(1)}_{2},\lambda^{(1)}_{1})$ have to be 
related in some sense. The commutation rule (29) itself gives us a hint 
how these two vectors can be related to each other. In fact, it is not 
difficult to conclude that the following exchange property 
\EQ
\vec{\Phi_2}(\lambda^{(1)}_{1},\lambda^{(1)}_{2}) = \vec{\Phi}_{2}(\lambda^{(1)}_{2},\lambda^{(1)}_{1}) . \frac{X^{(1)}(\lambda^{(1)}_{1}-\lambda^{(1)}_{2})}{
a(\lambda^{(1)}_{1} - \lambda^{(1)}_{2}) }
\EN
is valid, provided the function $\hat{h}(\lambda^{(1)}_{1},\lambda^{(1)}_{2})$ is 
fixed as in equation (42). To prove (43) we have used the remarkable 
identity
\EQ
\vec{\xi}. X^{(1)}(\lambda^{(1)}_{1}-\lambda^{(1)}_{2}) =
\frac{\hat{h}(\lambda^{(1)}_{1},\lambda^{(1)}_{2})}
{\hat{h}(\lambda^{(1)}_{2},\lambda^{(1)}_{1})} 
\{ 1 -\hat{t}b(\lambda^{(1)}_{1}-\lambda^{(1)}_{2}) \} \vec{\xi}
\EN
which relates vector $\vec{\xi}$, the auxiliary matrix $X^{(1)}$ and the 
Boltzmann weights. This is another way to see the role of vector $\vec{\xi}$;
it is an eigenvector of the auxiliary matrix $X^{(1)}$ with defined 
eigenvalue where the function $\hat{h}(\lambda^{(1)}_{1},\lambda^{(1)}_{2})$ 
enters in a symmetrical way. 

In order to cancel out other remaining `unwanted terms', it is 
necessary to impose further restriction on variables 
$\{\lambda^{(1)}_{1},\lambda^{(1)}_{2}\}$. In appendix $A$ we show how many 
different kinds of `unwanted terms' can be canceled out by the following 
Bethe Ansatz equations
\EQ
\left[ \frac{a(\lambda^{(1)}_{i})}{b(\lambda^{(1)}_{i})} \right]^{L}
\prod_{j=1 \; j \neq i}^{2} b(\lambda^{(1)}_{i}-\lambda^{(1)}_{j})
\frac{a(\lambda^{(1)}_{j}-\lambda^{(1)}_{i})}{b(\lambda^{(1)}_{j}-\lambda^{(1)}_{i})}
{\cal F}^{a_{2}a_{1}} =
{\tilde{X}}^{(1)}(\lambda^{(1)}_{i}-\lambda^{(1)}_{j}) _{a_{2}a_{1}}^{b_{1}b_{2}} {\cal F}^{b_{2}b_{1}} ,~~ i=1,2.
\EN

The same kind of reasoning can be used in order to construct the 3-particle 
state. Here, however, it is already interesting to write it in terms of certain 
recurrence structure. Inspired in our previous construction for the supersymmetric 
$spl(2|1)$ model, we begin with the following Ansatz
\bear
&& \vec{\Phi}_{3}(\lambda^{(1)}_{1},\lambda^{(1)}_{2},\lambda^{(1)}_{3}) = \vec{B}(\lambda^{(1)}_{1}) 
\otimes \vec{\Phi}_{2}(\lambda^{(1)}_{2},\lambda^{(1)}_{3}) + 
\nonumber \\
&& \left[ \vec{\xi} \otimes  F(\lambda^{(1)}_{1}) \vec{B}(\lambda^{(1)}_{3}) B(\lambda^{(1)}_{2}) \right]
 \hat{h}_{1}(\lambda^{(1)}_{1},\lambda^{(1)}_{2},\lambda^{(1)}_{3})   
+ \left[ \vec{\xi} \otimes  F(\lambda^{(1)}_{1}) \vec{B}(\lambda^{(1)}_{2}) B(\lambda^{(1)}_{3}) \right]    \hat{h}_{2}(\lambda^{(1)}_{1},\lambda^{(1)}_{2},\lambda^{(1)}_{3}) 
\nonumber \\
\ear

As before, the functions $\hat{h}_{1}(\lambda^{(1)}_{1},\lambda^{(1)}_{2},\lambda^{(1)}_{3})$ and 
$\hat{h}_{2}(\lambda^{(1)}_{1},\lambda^{(1)}_{2},\lambda^{(1)}_{3})$ can be fixed 
either by collecting together the `easy unwanted' terms or by using the permutation 
symmetries $\lambda^{(1)}_{2} \rightarrow \lambda^{(1)}_{3}$ and 
$\lambda^{(1)}_{1} \rightarrow \lambda^{(1)}_{2}$. We have found that they 
are given 
by
\bear
\hat{h}_{1}(\lambda^{(1)}_{1},\lambda^{(1)}_{2},\lambda^{(1)}_{3}) = 
- \frac{d_n(\lambda^{(1)}_{1}-\lambda^{(1)}_{2})}{e_n(\lambda^{(1)}_{1}-\lambda^{(1)}_{2})}  
 \frac{a(\lambda^{(1)}_{3}-\lambda^{(1)}_{2})}{b(\lambda^{(1)}_{3}-\lambda^{(1)}_{2})} I
\\
\hat{h}_{2}(\lambda^{(1)}_{1},\lambda^{(1)}_{2},\lambda^{(1)}_{3}) =
- \frac{d_n(\lambda^{(1)}_{1}-\lambda^{(1)}_{3})}{e_n(\lambda^{(1)}_{1}-\lambda^{(1)}_{3})}  
 \frac{1}{b(\lambda^{(1)}_{2}-\lambda^{(1)}_{3})} X_{23}^{(1)}(\lambda^{(1)}_{2}-\lambda^{(1)}_{3}) 
\ear

While the permutation $\lambda^{(1)}_{2} \rightarrow \lambda^{(1)}_{3}$ is 
easily verified, that concerning the symmetry $\lambda^{(1)}_{1} \rightarrow \lambda^{(1)}_{2}$ is quite cumbersome. One of the  main difficulty, for example,
is that we 
have also to commute $\vec{B}(\lambda^{(1)}_{1})$  and
$F(\lambda^{(1)}_{2})$. Therefore,
the commutation relations $(31-32)$ between the creation fields play an 
important role to disentangle the permutation $\lambda^{(1)}_{1} \rightarrow 
\lambda^{(1)}_{2}$. Besides that, certain additional properties between auxiliary 
matrix $X^{(1)}$, vector $\vec{\xi}$ and the field $\vec{B}(\lambda)$ are also 
necessary. For sake of completeness,
we have collected the details of the analysis of the
three-particle state in Appendix B. Considering 
the structures of the two and three-particle 
state (equations (40) and (46)), it  becomes
clear that the $m_1$-particle state can be generated by means of a recurrence relation. 
The general $m_1$-particle state can be obtained by using an induction procedure, and 
our final result is given by\footnote{The indices under  
$X_{k,k+1}^{(1)}(\lambda^{(1)}_{k}-\lambda^{(1)}_{j})$ indicate the positions on the
 Hilbert space where this matrix acts in a non-trivial way.} 
%
\begin{eqnarray}
&& \vec {\Phi}_{m_{1}}(\lambda^{(1)}_{1},\lambda^{(1)}_{2}, \dots,  \lambda^{(1)}_{m_{1}})=
\nonumber \\
&& \vec {B}(\lambda^{(1)}_{1}) \otimes \vec {\Phi}_{n-1}(\lambda^{(1)}_{2}, \dots,  
 \lambda^{(1)}_{m_{1}})
-\sum_{j=2}^{m_{1}} 
\frac{d_n(\lambda^{(1)}_{1}-\lambda^{(1)}_{j})}{e_n(\lambda^{(1)}_{1}-\lambda^{(1)}_{j})} 
\prod_{k=2,k \neq j}^{m_{1}} 
\frac{a(\lambda^{(1)}_{k}-\lambda^{(1)}_{j})}{b(\lambda^{(1)}_{k}-\lambda^{(1)}_{j})} \nonumber \\
&& \times \left[ \vec{\xi} \otimes  F(\lambda^{(1)}_{1}) \vec {\Phi}_{n-2}(\lambda^{(1)}_{2},\dots,\lambda^{(1)}_{j-1},\lambda^{(1)}_{j+1},\dots, 
\lambda^{(1)}_{m_{1}} ) B(\lambda^{(1)}_{j}) \right] \prod_{k=2}^{j-1} 
\frac{X_{k,k+1}^{(1)}(\lambda^{(1)}_{k}-\lambda^{(1)}_{j})}{
a(\lambda^{(1)}_{k} - \lambda^{(1)}_{j})}  
\nonumber \\
\end{eqnarray}
%
and under a consecutive permutation  $\lambda^{(1)}_{j} \rightarrow \lambda^{(1)}_{j+1}$
 the $m_1$-particle state satisfies the following relation
\EQ
\vec {\Phi}_{m_{1}}(\lambda^{(1)}_{1},\cdots,\lambda^{(1)}_{j},\lambda^{(1)}_{j+1},\cdots, \lambda^{(1)}_{m_{1}}) = 
\vec {\Phi}_{m_{1}}(\lambda^{(1)}_{1},\cdots,\lambda^{(1)}_{j+1},\lambda^{(1)}_{j},\cdots, \lambda^{(1)}_{m_{1}}) 
\frac{X_{j,j+1}^{(1)}(\lambda^{(1)}_{j}-\lambda^{(1)}_{j+1})}{a(\lambda^{(1)}_{j}-\lambda^{(1)}_{j+1})}
\EN

The following remark is now in order. So far we have not commented about the role of 
the creation field $\vec{B^{*}}(\lambda)$. From the integrability condition (14), we can 
verify that the commutation 
rules between the creation field $\vec{B^{*}}(\lambda)$ with 
the diagonal operators and with 
the scalar creation field $F(\lambda)$ are similar to  those
in equations $(22-24,26-32)$. Basically, we have 
to change $\vec{B}(\lambda)$ by 
$\vec{B^{*}}(\lambda)$, to replace the definitions of $\hat{A}(\lambda)$ and $\vec{\xi}$ by 
their transpose ones, and to interchange the diagonals operators
$B(\lambda) \leftrightarrow D(\lambda)$. The three creation fields do not mix all together, but only in pair as 
$\{ \vec{B^{*}}(\lambda),F(\lambda) \}$ and 
$ \{ \vec{B}(\lambda),F(\lambda) \}$. We shall then call  
$\vec{B^{*}}(\lambda)$ the `` dual '' of $\vec{B}(\lambda)$. This
dual creation field  together with $F(\lambda)$
can also be used to built eigenvectors, and by using the replacements
mentioned above we obtain an 
expression analogous to that of equation (49). We think that these two
possible ways of constructing the eigenvectors go back to the fact that the
vertex operator ${\cal{L}}_{{\cal{A}}i}(\lambda)$ becomes naturally crossing
invariant after multiplying it by the factor $(\Delta-\lambda)$. Indeed,
it is not difficult to see that their corresponding eigenvalues are 
related to each other by performing the crossing shift $x \rightarrow \Delta -x$
both in the transfer matrix parameter and in all the Bethe ansatz rapidities.
It seems 
interesting to analyse to what extent we 
can benefit from this property of the eigenvectors in order to  provide
new physical insights for these vertex models.

Let us now turn to the Bethe Ansatz conditions on the variables $\{\lambda^{(1)}_{i}\}$.
 A typical unwanted term, coming from the eigenvalue problem 
$(18)$, arises when the variables $\lambda^{(1)}_{i}$ of the 
$m_1$-particle state is exchanged 
with the transfer matrix rapidity $\lambda$. For instance, this is the case of an unwanted 
term having the structure $B_{a_{1}}(\lambda)B_{a_2}(\lambda^{(1)}_{2}) \cdots B_{a_{m_{1}}}( \lambda^{(1)}_{m_{1}})$. This unwanted term is produced by the action of the diagonal 
operators $B(\lambda)$ and $\sum_{a} \hat{A}_{aa}(\lambda)$ on the 
$m_1$-particle state.
From the commutation rules $(22,23)$ 
we are able to show that they can be collected in the 
following forms
\EQ
-\frac{[a(\lambda^{(1)}_{1})]^{L}}
{b(\lambda^{(1)}_{1}-\lambda)}
\prod_{j=2}^{m_{1}} \frac{a(\lambda^{(1)}_{j}-\lambda^{(1)}_{1})}
{b(\lambda^{(1)}_{j}-\lambda^{(1)}_{1})} 
{\cal F}^{a_{m_{1}} \cdots a_{1}}
B_{a_{1}}(\lambda)B_{a_{2}}(\lambda^{(1)}_{2}) \cdots B_{a_{m_{1}}}( \lambda^{(1)}_{m_{1}}) \ket{0}
\EN
and
\EQ
\frac{[b(\lambda^{(1)}_{1})]^{L} }
{b(\lambda^{(1)}_{1}-\lambda)}
\prod_{j=2}^{m_{1}} \frac{1}{b(\lambda^{(1)}_{1}-\lambda^{(1)}_{j})}
T^{(1)}(\lambda=\lambda^{(1)}_{1},\{\lambda^{(1)}_{i}\})_{b_{1} \cdots b_{m_{1}}}^{a_{1} \cdots a_{m_{1}}} {\cal F}^{a_{m_{1}} \cdots a_{1}}
B_{b_{1}}(\lambda) B_{b_{2}}(\lambda^{(1)}_{2}) 
\cdots B_{b_{m_{1}}}( \lambda^{(1)}_{m_{1}}) \ket{0}
\EN
%
where $T^{(1)}(\lambda,\{\lambda^{(1)}_{i}\})$ is the transfer matrix of a 
inhomogeneous vertex model defined in terms of 
the auxiliary $R$-matrix ${\tilde{X}}^{(1)}(\lambda)$ by
\EQ
T^{(1)}(\lambda,\{\lambda^{(1)}_{i}\})_{b_{1} \cdots b_{m_{1}}}^{a_{1} \cdots a_{m_{1}}} = 
[{\tilde{X}}^{(1)}(\lambda-\lambda^{(1)}_1)]_{b_{1}d_{1}}^{c_{1}a_{1}}
[{\tilde{X}}^{(1)}(\lambda-\lambda^{(1)}_2)]_{b_{2}c_{2}}^{d_{1}a_{2}} \cdots
[{\tilde{X}}^{(1)}(\lambda-\lambda^{(1)}_{m_1})]_{b_{m_{1}}c_{1}}^{d_{n-1}a_{m_{1}}}
\EN

Similarly, the same kind of reasoning can be pursued for any $\lambda^{(1)}_{i}$, 
thanks to the property (50) which relates vector 	
$\vec{\Phi}_{m_{1}}(\lambda^{(1)}_{1},\cdots, \lambda^{(1)}_{m_{1}})$ to 
$\vec{\Phi}_{m_{1}}(\lambda^{(1)}_{i},\cdots, 
\lambda^{(1)}_{m_{1}})$  by cyclic permutations. Thus, 
from equations (51) and (52) the unwanted term 
$B_{a_{1}}(\lambda^{(1)}_{1}) \dots B_{a_{i}}(\lambda) \cdots B_{a_{m_{1}}}( \lambda^{(1)}_{m_{1}})$ can be canceled out provided that 
\bear
&& \left[ \frac{a(\lambda^{(1)}_{i})}{b(\lambda^{(1)}_{i})} \right]^{L}
\prod_{j=1 \; j \neq i}^{m_{1}} b(\lambda^{(1)}_{i}-\lambda^{(1)}_{j})
\frac{a(\lambda^{(1)}_{j}-\lambda^{(1)}_{i})}{b(\lambda^{(1)}_{j}-\lambda^{(1)}_{i})}
{\cal F}^{a_{m_{1}} \dots a_{1}} =
\nonumber \\
&& T^{(1)}(\lambda=\lambda^{(1)}_{i},\{\lambda^{(1)}_{j}\})_{b_{1} \cdots b_{m_{1}}}^{a_{1} \cdots a_{m_{1}}} {\cal F}^{b_{m_{1}} \dots b_{1}} , i=1, \dots, m_{1}
\ear

This generalizes equation $(45)$ for an arbitrary value of the number of 
`particles' $m_{1}$.
 In general, the action of the diagonal operators on the $m_{1}$-particle state 
generates many other species of unwanted terms. A systematic way to collect all of them in 
families of unwanted terms with a 
defined structure for a general value of $m_{1}$ has eluded us 
so far. However, we remark that for those we have been able to catalog,
such as
$B^{*}_{a_{1}}(\lambda^{(1)}_{1}) 
\dots B_{a_{i}}(\lambda) \cdots B_{a_{m_{1}}}( \lambda^{(1)}_{m_{1}})$,
condition (54) has been explicitly verified. 
Furthermore, the explicit checks we have performed in two and three-particle states 
leaves no doubt that restriction (54) is the unique condition\footnote{In a factorizable 
theory, which is our case here, the two-particle structure already contains the main flavour 
about the Bethe Ansatz equations ( see equation (45) ). } 
to be imposed on the variables $\{\lambda^{(1)}_{1},\cdots, \lambda^{(1)}_{m_{1}}\}$. 
Anyhow, a rigorous proof should be welcome, because probably it 
will shed extra light to the 
mathematical structure we have found for the eigenvectors. 

By the same token, the eigenvalue $\Lambda(\lambda,\{\lambda^{(1)}_{i}\})$ of the $m_{1}$-particle state can be calculated by keeping only terms proportional to the eigenstate 
$\ket{\Phi_{m_{1}}(\lambda^{(1)}_{1},\cdots, \lambda^{(1)}_{m_{1}})}$. For instance, 
that proportional to the term 
$B_{a_{1}}(\lambda^{(1)}_{1}) \dots B_{a_{i}}(\lambda^{(1)}_{i}) \cdots B_{a_{m_{1}}}( \lambda^{(1)}_{m_{1}})$ can be determined by the commutation relations 
$(22-24)$. By keeping the first terms each time we turn the diagonal operators 
over one of the $\vec{B}(\lambda)$ components, we find that
\bear
\Lambda(\lambda,\{\lambda^{(1)}_{i}\}) = &&
[a(\lambda)]^L \prod_{i=1}^{m_{1}} \frac{a(\lambda^{(1)}_{i}-\lambda)}{b(\lambda^{(1)}_{i}-\lambda)} +
[e_{n}(\lambda)]^L \prod_{i=1}^{m_{1}} \frac{b(\lambda-\lambda^{(1)}_{i})}{e_{n}(\lambda-\lambda^{(1)}_{i})} +
\nonumber \\
&& [b(\lambda)]^L \prod_{i=1}^{m_{1}}
\frac{1}{b(\lambda-\lambda^{(1)}_{i})}
\Lambda^{(1)}(\lambda,\{\lambda^{(1)}_{i}\})  
\ear
where $\Lambda^{(1)}(\lambda,\{\lambda^{(1)}_{i}\})$ is the eigenvalue of the 
inhomogeneous transfer-matrix $T^{(1)}(\lambda,\{\lambda^{(1)}_{i}\})$, 
\EQ
T^{(1)}(\lambda,\{\lambda^{(1)}_{i}\})_{b_{1} \cdots b_{m_{1}}}^{a_{1} \cdots a_{m_{1}}} {\cal F}^{b_{m_{1}} \dots b_{1}} =
\Lambda^{(1)}(\lambda,\{\lambda^{(1)}_{i}\}) {\cal F}^{a_{m_{1}} \dots a_{1}}
\EN
and therefore the Bethe Ansatz equations (54) are then disentangled in terms of 
the eigenvalue $\Lambda^{(1)}(\lambda,\{\lambda^{(1)}_{i}\})$ as
\EQ
\left[ \frac{a(\lambda^{(1)}_{i})}{b(\lambda^{(1)}_{i})} \right]^{L}
\prod_{j=1 \; j \neq i}^{m_{1}} b(\lambda^{(1)}_{i}-\lambda^{(1)}_{j})
\frac{a(\lambda^{(1)}_{j}-\lambda^{(1)}_{i})}{b(\lambda^{(1)}_{j}-\lambda^{(1)}_{i})} =
\Lambda^{(1)}(\lambda = \lambda^{(1)}_{i},\{\lambda^{(1)}_{j}\})
,~~ i=1, \dots, m_{1}
\EN

This completes the first step of our analysis, because we still need to 
find the eigenvalue $\Lambda^{(1)}(\lambda,\{\lambda^{(1)}_{i}\})$ of the 
transfer-matrix $T^{(1)}(\lambda,\{\lambda^{(1)}_{i}\})$. This is the 
so-called nested Bethe Ansatz problem and we shall deal with it in next 
section.
\section{The nested Bethe Ansatz problem}

In the last section we were left with the problem of diagonalization of 
the inhomogeneous transfer matrix $T^{(1)}(\lambda,\{\lambda^{(1)}_{i}\})$. 
This problem can still be solved by the quantum inverse scattering approach, 
once the Yang-Baxter algebra (14) is extended to accommodate the presence of 
inhomogeneities. Formally, the corresponding monodromy matrix can be written as 
\EQ
{\cal{T}}^{(1)}(\lambda,\{\lambda^{(1)}_{i}\}) =
{\cal L}^{(1)}_{{\cal A}m_{1}}(\lambda-\lambda^{(1)}_{m_{1}}){\cal L}^{(1)}_{{\cal A}m_{1}-1}
(\lambda-\lambda^{(1)}_{m_{1}-1}) \cdots {\cal L}^{(1)}_{{\cal A}1}(\lambda-\lambda^{(1)}_{1})
\EN
where the components of the vertex operator $[{\cal L}^{(1)}(\lambda)]_{ab}^{cd}$ are 
related to that of the $R$-matrix 
$[{\tilde{X}}^{(1)}(\lambda)]_{ab}^{cd}$ by the standard 
permutation
\EQ
\left[{\cal L}^{(1)}(\lambda) \right]_{ab}^{cd} = 
\left[ {\tilde{X}}^{(1)}(\lambda) \right]_{ba}^{cd}
\EN

In order to go on, we need first to analyse the properties of the 
auxiliary matrix 
${\tilde{X}}^{(1)}(\lambda)$. For instance, we have to verify if the 
triangular property of 
${\cal L}^{(1)}(\lambda)$ still has the same structure as 
that present on the original vertex operator ${\cal L}(\lambda)$ we 
started with
( see equation (16) ). In other words, we have to check 
whether or not the basic 
structure of the Ansatz (17) for the monodromy matrix is still appropriate. 
As we shall see, this depends much on the type of the original vertex model we 
are diagonalizing and also on their number of states per link $q$. We recall 
that in the first step we already lost two degrees of freedom, remaining 
$(q-2)$ states per link to represent the Boltzmann weights of matrix
${\tilde{X}}^{(1)}(\lambda)$. Hence, if $(q-2) < 3$ 
the Ansatz (17) certainly is not more the convenient one.

Let us, for the moment, suppose that the $R$-matrix 
${\tilde{X}}^{(1)}(\lambda)$
 keeps the basic properties of the original model. In this case, we have 
basically to adapt our main results of sections 3 and 4 to include the 
inhomogeneities $\{\lambda^{(1)}_{1}, \dots, \lambda^{(1)}_{m_{1}}\}$. For 
instance, the pseudovacuum $\ket{0^{(1)}}$ is the usual ferromagnetic 
state, but now with length $(q-2)$
\EQ
\ket{0^{(1)}} = \prod_{i=1}^{m_{1}} \otimes \ket{0^{(1)}}_{i} , ~~
\ket{0^{(1)}}_{i} = 
\pmatrix{
1 \cr
0 \cr
\vdots \cr
0 \cr}_{q-2}
\EN
and the vertex operator ${\cal L}^{(1)}_{{\cal A}i}(\lambda-\lambda^{(1)}_{i})$ has 
the following triangular property
{\scriptsize
\EQ
{\cal L}^{(1)}_{{\cal A}i}(\lambda-\lambda^{(1)}_{i})\ket{0^{(1)}}_{i} = 
\pmatrix{
a(\lambda-\lambda^{(1)}_{i}) \ket{0^{(1)}}_i &  *  &  *  & \dots & * & *  \cr
0  &  b(\lambda-\lambda^{(1)}_{i}) \ket{0^{(1)}}_i &  0  & \dots & 0 & *  \cr
\vdots & 0 & \ddots &  & 0 & \vdots \cr
\vdots & \vdots &  & \ddots & 0 & \vdots \cr
0  &  0  &  0  & \dots & b(\lambda-\lambda^{(1)}_{i}) \ket{0^{(1)}}_i & * \cr
0  &  0  &  0  & \dots & 0 &  e_{n-1}(\lambda-\lambda^{(1)}_{i}) \ket{0^{(1)}}_i
 \cr}_{(q-2) \times (q-2)} 
\EN
}

As before, if we assume the following structure for the monodromy matrix 
${\cal T}^{(1)}(\lambda,\{\lambda^{(1)}_{i}\})$ 
\EQ
{\cal T}^{(1)}(\lambda,\{\lambda^{(1)}_{i}\}) = 
\pmatrix{
B^{(1)}(\lambda,\{\lambda^{(1)}_{i}\})       &   \vec{B}^{(1)}(\lambda,\{\lambda^{(1)}_{i}\})   &   F^{(1)}(\lambda,\{\lambda^{(1)}_{i}\})   \cr
\vec{C}^{(1)}(\lambda,\{\lambda^{(1)}_{i}\})  &  \hat{A}^{(1)}(\lambda,\{\lambda^{(1)}_{i}\})   &  \vec{B^{*}}^{(1)}(\lambda,\{\lambda^{(1)}_{i}\})   \cr
C^{(1)}(\lambda,\{\lambda^{(1)}_{i}\})  & \vec{C^{*}}^{(1)}(\lambda,\{\lambda^{(1)}_{i}\})  &  D^{(1)}(\lambda,\{\lambda^{(1)}_{i}\})  \cr}_{(q-2) \times (q-2)}
\EN
we find that the operators of the monodromy matrix 
${\cal T}^{(1)}(\lambda,\{\lambda^{(1)}_{i}\})$ satisfy the following `diagonal' properties 
\bear
B^{(1)}(\lambda,\{\lambda^{(1)}_{i}\})\ket{0^{(1)}} = \prod_{i=1}^{m_{1}} a(\lambda-\lambda^{(1)}_{i})\ket{0^{(1)}};~~
D^{(1)}(\lambda,\{\lambda^{(1)}_{i}\})\ket{0^{(1)}} =  \prod_{i=1}^{m_{1}}
e_{n-1}(\lambda-\lambda^{(1)}_{i})\ket{0^{(1)}};
\nonumber \\ 
A^{(1)}_{aa}(\lambda,\{\lambda^{(1)}_{i}\})\ket{0^{(1)}} =  \prod_{i=1}^{m_{1}} b(\lambda-\lambda^{(1)}_{i})\ket{0^{(1)}} ,~~ a=1, \dots , q-4
\ear
as well as the annihilation properties
\bear
C^{(1)}_{a}(\lambda,\{\lambda^{(1)}_{i}\})\ket{0^{(1)}} = 0;~~ 
C_{a}^{*(1)}(\lambda,\{\lambda^{(1)}_{i}\})\ket{0^{(1)}} = 0; 
\nonumber \\
C^{(1)}(\lambda,\{\lambda^{(1)}_{i}\}) \ket{0^{(1)}} = 0; ~~ 
A^{(1)}_{ab}(\lambda,\{\lambda^{(1)}_{i}\})\ket{0^{(1)}} = 0,  ~(a,b =1, \cdots , q-4; a \neq b)
\ear

In order to diagonalize the transfer matrix 
 $T^{(1)}(\lambda,\{\lambda^{(1)}_{i}\})$ we have to introduce a  new set 
of variables $\{\lambda^{(2)}_{1}, \dots, \lambda^{(2)}_{m_{2}}\}$,
which are going to parametrize  the 
multi-particle states 
of the nesting problem at step $1$. Evidently, the structure of the 
commutation rules $(22-24;26-32)$ and 
the eigenvector construction of section 4 
remains precisely the same. We basically have to change $q$ to $(q-2)$ in the 
Boltzmann weights, a given operator $\hat{O}(\lambda)$ by its 
corresponding  $\hat{O}^{(1)}(\lambda,\{\lambda^{(1)}_{i}\})$ and to 
replace the parameters $\{\lambda^{(1)}_{1}, \dots, \lambda^{(1)}_{m_{1}}\}$
by $\{\lambda^{(2)}_{1}, \dots, \lambda^{(2)}_{m_{2}}\}$ in the eigenvector 
expression (49). It turns out that the eigenvalue expression 
$\Lambda^{(1)}(\lambda,\{\lambda^{(1)}_{i}\})$ and the corresponding Bethe 
Ansatz equations will again depend on  another auxiliary inhomogeneous 
vertex model, having now $(q-4)$ states per link. Of course, we can repeat this 
procedure until we reach a certain  
step $l$ where the underlying auxiliary $R$-matrix 
${\tilde{X}}^{(l)}$ lost the 
basic features presented by the original $R$-matrix  we 
began with. In general, by using this ``nested'' procedure we can find a 
relation between the eigenvalues 
$\Lambda^{(l)}(\lambda,\{\lambda^{(l)}_{j}\}, 
\dots ,\{\lambda^{(1)}_{j}\})$ and 
$\Lambda^{(l+1)}(\lambda,\{\lambda^{(l+1)}_{j}\}, \dots ,\{\lambda^{(1)}_{j}\})$, on 
the steps $l$ and $l+1$, respectively. We basically 
have to `` dress '' our previous result (55)
with inhomogeneities, and to consider the appropriate Boltzmann 
weights of the step $l$ we are diagonalizing. This relation can be written as 
\bear
&&\Lambda^{(l)} \left(\lambda;\{\lambda^{(l)}_j\}, \cdots,\{\lambda^{(1)}_j\} \right) =
\nonumber \\
&& \prod_{j=1}^{m_{l}} a(\lambda-\lambda^{(l)}_{j}) \prod_{j=1}^{m_{l+1}} 
\frac{a(\lambda^{(l+1)}_{j} - \lambda)}{b(\lambda^{(l+1)}_{j} - \lambda)} 
+\prod_{j=1}^{m_{l}} e_{(n-l)}(\lambda-\lambda^{(l)}_{j}) 
\prod_{j=1}^{m_{l+1}} \frac{b(\lambda-\lambda^{(l+1)}_{j})}{e_{(n-l)}
(\lambda-\lambda^{(l+1)}_{j})} +
\nonumber \\
&& \prod_{j=1}^{m_{l}} b(\lambda-\lambda^{(l)}_{j}) \prod_{j=1}^{m_{l+1}} \frac{1}{b(\lambda-\lambda^{(l+1)}_{j})} 
\Lambda^{(l+1)} \left(\lambda;\{\lambda^{(l+1)}_j \}, 
\{\lambda^{l}_j \}, \cdots,\{\lambda^{(1)}_j \} \right)
\ear
This last equation has to be understood as a 
recurrence relation, beginning on step zero.
In order to be consistent, we 
identify the zero step $l=0$ with the eigenvalue of the 
original transfer matrix $T(\lambda)$ we wish to 
diagonalize. Therefore, we are assuming the following 
identifications $\Lambda^{(0)}=\Lambda$, 
$\{\lambda^{(0)}_{i}\} \equiv {0}$ and $m_{0} \equiv L$. 
Analogously, the Bethe Ansatz restriction on the variables 
$\{\lambda^{(l+1)}_{j}\}$ introduced to parametrize the Fock 
space of the inhomogeneous transfer matrix $T^{(l)}(\lambda, 
\{\lambda^{(l)}_{j}\}, \dots ,\{\lambda^{(1)}_{j}\})$ 
on step $l$ is given by
\bear
&& \prod_{i=1}^{m_{l}} \frac{a(\lambda^{(l+1)}_{j}-\lambda^{(l)}_{i})}{b(\lambda^{(l+1)}_{j}-\lambda^{(l)}_{i})}
\prod_{i=1 \; i \neq j}^{m_{l+1}} b(\lambda^{(l+1)}_{j}-\lambda^{(l+1)}_{i})
\frac{a(\lambda^{(l+1)}_{i}-\lambda^{(l+1)}_{j})}{b(\lambda^{(l+1)}_{i}-\lambda^{(l+1)}_{j})}
 =
\nonumber \\
&& \Lambda^{(l+1)}(\lambda=\lambda^{(l+1)}_{j},\{\lambda^{(l+1)}_{j}\}, \cdots ,\{\lambda^{(1)}_{j}\}),~~ j=1, \cdots, m_{l+1}
\ear

We now will particularize our discussion concerning the nesting structure
for each pair of models $\{ B_n, Osp(2n-1|2) \} $, $\{ C_n, Osp(2|2n-2) \}$,
$ \{ D_n, Osp(2n-2|2) \} $ and for the $Osp(1|2n)$ vertex models. For the pair 
$\{ B_n, Osp(2n-1|2) \} $, it is not difficult to check that the nesting
structure developed above ( see equations (65,66) ) works
in any step. In this case, the last step consists of 
an inhomogeneous $B_{1}$ vertex 
model and the respective  $R$-matrix acts on  $C^{3} \times C^{3}$ 
tensor space. The vertex 
operator ${\cal L}^{B_{1}}(\lambda)$  has the following structure
\EQ
{\footnotesize
{\cal L}^{B_{1}}(\lambda) =
  \pmatrix{
 a(\lambda)  &0  &0  &0  &0  &0  &0  &0  &0   \cr
 0  &b(\lambda)  &0  &1  &0  &0  &0  &0  &0     \cr
 0  &0  &e_{1}(\lambda)  &0  &d_{1}(\lambda)  &0  &c_{1}(\lambda)  &0  &0    \cr
 0  &1  &0  &b(\lambda)  &0  &0  &0  &0  &0     \cr
 0  &0  &d_{1}(\lambda)  &0  &f_{1}(\lambda)  &0  &d_{1}(\lambda)  &0  &0    \cr
 0  &0  &0  &0  &0  &b(\lambda)  &0  &1  &0     \cr
 0  &0  &c_{1}(\lambda)  &0  &d_{1}(\lambda)  &0  &e_{1}(\lambda)  &0  &0    \cr
 0  &0  &0  &0  &0  &1  &0  &b(\lambda)  &0     \cr
 0  &0  &0  &0  &0  &0  &0  &0  &a(\lambda)   \cr}
}
\EN
where the Boltzmann weights $a(\lambda)$, $b(\lambda)$, $c_1(\lambda)$,
$d_1(\lambda)$ and $e_1(\lambda)$ are listed on table 2 for $B_1$, and 
$f_1(\lambda)=b(\lambda)+c_1(\lambda)$.
Due to the isomorphism $B_1 \equiv O(3) \sim SU(2)_{k=2}$, this vertex 
model is 
equivalent to the isotropic spin-$1$ XXX model \cite{TABA}. Therefore, 
the $B_1$ model can be solved 
either by adapting the known results for 
the spin-$1$ XXX model \cite{TABA} to include 
inhomogeneities or by applying the general approach we have developed in 
sections 3 and 4. In the latter case, we remark that our construction for the
eigenvalues, eigenvectors and commutation rules reduce to that 
proposed prviously
by Tarasov in the context of the Izergin-Korepin vertex model \cite{TA}.
The final result we have found for the 
eigenvalue $\Lambda^{B_{1}}(\lambda,\{\lambda_{i}\},\{\mu_{j}\})$ of the
inhomogeneous $B_1$ vertex model is  
\bear
&& \Lambda^{B_{1}}\left(\lambda,\{\lambda_{i}\},\{\mu_{j}\}\right) =
 \prod_{i=1}^{n} (\lambda-\lambda_{i}+1) \prod_{j=1}^{m} \frac{\mu_{j}-\lambda+1}{\mu_{j}-\lambda} +
\nonumber \\
&& \prod_{i=1}^{n} (\lambda-\lambda_{i}) \prod_{j=1}^{m}  \frac{(\lambda-\mu_{j}+1/2)}{(\lambda-\mu_{j})} \frac{(\lambda-\mu_{j}-1)}{(\lambda-\mu_{j}-1/2)}+ 
\prod_{i=1}^{n} \frac{(\lambda-\lambda_{i})(\lambda-\lambda_{i}-1/2)}{(\lambda-\lambda_{i}+1/2)}
\prod_{j=1}^{m}  \frac{\lambda-\mu_{j}+1/2} {\lambda-\mu_{j}-1/2}
\nonumber \\
\ear
where $\{ \lambda_1, \cdots, \lambda_n \} $ are the
inhomogeneities and variables $\{\mu_{j}\}$ satisfy the following 
Bethe Ansatz equation
\EQ
\prod_{i=1}^{n} \frac{\mu_{j}-\lambda_{i}+1}{\mu_{j}-\lambda_{i}} = 
 \prod_{k=1 \; k \neq j}^{m} \frac{\mu_{j}-\mu_{k}+1/2}{\mu_{j}-\mu_{k}-1/2} , ~~ j=1, \dots,m.
\EN

On the other hand, the situation for the pairs of models $
 \{ C_{n}$, $Osp(2|2n-2) \} $ and 
$ \{D_{n}$, $Osp(2n-2|2) \}$ is a little bit different. In these 
cases, we can proceed by using the 
recurrence relation (65) and the Bethe Ansatz condition (66) until we reach the steps 
$l=n-1$ and $l=n-2$, respectively. For the $C_{n}$ and $Osp(2|2n-2)$ 
vertex models the  nesting level
$l=n-1$ corresponds to the diagonalization of an inhomogeneous transfer matrix 
which Boltzmann 
weights  have the $6$-vertex symmetry. Indeed,  at this level we have to
diagonalize 
the inhomogeneous $C_{1}$ system whose vertex operator
${\cal L}^{C_{1}}(\lambda)$ is given by
\EQ
{\cal L}^{C_{1}}(\lambda) = 
(\lambda+1)
\pmatrix{
1  &0  &0  &0  \cr
0  &\frac{\lambda}{\lambda+2}  &\frac{2}{\lambda+2}  &0  \cr
0  &\frac{2}{\lambda+2}  &\frac{\lambda}{\lambda+2}  &0  \cr
0  &0  &0  &1  \cr}
\EN

The diagonalization of the $6$-vertex models 
on a irregular lattice has appeared in many 
different context in the literature (see for instance refs. \cite{BA,KRA,KV,PM,KE}). By 
adapting these results, in order to consider the particular structure of ${\cal L}^{C_{1}}(\lambda)$,
we find that the eigenvalue $\Lambda^{C_{1}}\left(\lambda,\{\lambda_{i}\},\{\mu_{j}\}
\right)$ is 
\EQ
\Lambda^{C_{1}}\left(\lambda,\{\lambda_{i}\},\{\mu_{j}\} \right) =
\prod_{i=1}^{n} (\lambda-\lambda_{i}+1) \prod_{j=1}^{m} \frac{\mu_{j}-\lambda+2}{\mu_{j}-\lambda} +
\prod_{i=1}^{n} \frac{(\lambda-\lambda_{i})(\lambda-\lambda_{i}+1)}{(\lambda-\lambda_{i}+2)} \prod_{j=1}^{m}  \frac{\lambda-\mu_{j}+2} {\lambda-\mu_{j}}
\nonumber \\
\EN
where variables $\{\mu_{j}\}$ satisfy the Bethe Ansatz condition
\EQ
\prod_{i=1}^{n} \frac{\mu_{j}-\lambda_{i}}{\mu_{j}-\lambda_{i}+2} = 
\prod_{k=1 \; k \neq j}^{m} \frac{\mu_{j}-\mu_{k}-2}{\mu_{j}-\mu_{k}+2} , ~~ j=1, \dots,m.
\EN

We now turn to the analysis of the last nesting level $l=n-2$  
for the models $D_{n}$ and $Osp(2n-2|2)$. The final  
stage for these systems consists  of the diagonalization of the $D_{2}$ vertex 
model\footnote{Naively, one would think that this model can still be reduced to the $D_{1}$ vertex
 model. This is not the case, because the $R$-matrix of the the $D_{1}$ vertex model 
($\Delta=0$ in equation (4)) is no longer  regular at $\lambda=0$. }. It turns
out, however, that the 
$D_{2}$ weights can be decomposed in terms of the tensor product of two 
$6$-vertex models. More precisely, the vertex operator ${\cal L}^{D_{2}}(\lambda)$ can be 
written as 
\EQ
{\cal L}^{D_{2}}(\lambda) = 
{\cal L}^{6-ver}_{\sigma} \otimes {\cal L}^{6-ver}_{\tau}
\EN
where the two $6$-vertex structures are given by
\bear
{\cal L}^{6-ver}_{\sigma}(\lambda) = 
\pmatrix{
1  &0  &0  &0  \cr
0  &\frac{\lambda}{\lambda+1}  &\frac{1}{\lambda+1}  &0  \cr
0  &\frac{1}{\lambda+1}  &\frac{\lambda}{\lambda+1}  &0  \cr
0  &0  &0  &1  \cr}, ~~
{\cal L}^{6-ver}_{\tau}(\lambda) = 
\pmatrix{
\lambda+1  &0  &0  &0  \cr
0  &\lambda  &1  &0  \cr
0  &1  &\lambda  &0  \cr
0  &0  &0  &\lambda+1  \cr}
\ear

Consequently, the eigenvalues of the  model
$D_{2}$ are given in terms of the product of the 
eigenvalues 
of the two $6$-vertex models defined in equation (74). 
In the presence of  inhomogeneities 
$\{\lambda_{1}, \cdots, \lambda_{n}\}$ we find that 
these eigenvalues are given by
\bear
&& \Lambda^{D_{2}}\left(\lambda,\{\lambda_{i}\},\{\mu^{\pm}_{j}\}\right) =
\left[ 
\prod_{j=1}^{m_{+}}\frac{\lambda-\mu^{+}_{j}-1}{\lambda-\mu^{+}_{j}}  +
\prod_{i=1}^{n} \frac{\lambda-\lambda_{i}}{\lambda-\lambda_{i}+1} \prod_{j=1}^{m_{+}}  \frac{\lambda-\mu^{+}_{j}+1}{\lambda-\mu^{+}_{j}} \right] \times
\nonumber \\
&& \left[ 
\prod_{j=1}^{m_{-}}\frac{\lambda-\mu^{-}_{j}-1}{\lambda-\mu^{-}_{j}}\prod_{i=1}^{n} (\lambda-\lambda_{i}+1)  +
\prod_{j=1}^{m_{-}}  \frac{\lambda-\mu^{-}_{j}+1}{\lambda-\mu^{-}_{j}} \prod_{i=1}^{n} (\lambda-\lambda_{i}) \right]
\ear
where variables $\{\mu^{+}_{j}\}$ and
$\{\mu^{-}_{j}\}$ parametrize the multi-particle states of the inhomogeneous 
models related to 
${\cal L}^{6-ver}_{\sigma}(\lambda) $ and
${\cal L}^{6-ver}_{\tau}(\lambda) $, respectively. They satisfy the following
Bethe Ansatz equations
\EQ
\prod_{i=1}^{n} \frac{\mu^{\pm}_{j}-\lambda_{i}}{\mu^{\pm}_{j}-\lambda_{i}+1} = 
\prod_{k=1 \; k \neq j}^{m_{\pm}} \frac{\mu^{\pm}_{j}-\mu^{\pm}_{k}-1}{\mu^{\pm}_{j}-\mu^{\pm}_{k}+1} , ~~ j=1, \dots,m_{\pm}.
\EN

Finally, it remains to investigate the $Osp(1|2n)$ vertex model. In this case,
the nesting recurrence relations (65,66) are valid for any level. On the last
step  we have to deal with the inhomogeneous $Osp(1|2)$ model, possessing the
following  
vertex operator 
\EQ
{\footnotesize
{\cal L}^{Osp(1|2)}(\lambda) =
  \pmatrix{
 a(\lambda)  &0  &0  &0  &0  &0  &0  &0  &0   \cr
 0  &b(\lambda)  &0  &1  &0  &0  &0  &0  &0     \cr
 0  &0  &e_{1}(\lambda)  &0  &-d_{1}(\lambda)  &0  &c_{1}(\lambda)  &0  &0    \cr
 0  &1  &0  &b(\lambda)  &0  &0  &0  &0  &0     \cr
 0  &0  &-d_{1}(\lambda)  &0  &f_{1}(\lambda)  &0  &d_{1}(\lambda)  &0  &0    \cr
 0  &0  &0  &0  &0  &b(\lambda)  &0  &1  &0     \cr
 0  &0  &c_{1}(\lambda)  &0  &d_{1}(\lambda)  &0  &e_{1}(\lambda)  &0  &0    \cr
 0  &0  &0  &0  &0  &1  &0  &b(\lambda)  &0     \cr
 0  &0  &0  &0  &0  &0  &0  &0  &a(\lambda)   \cr}
}
\EN
where the Boltzmann weights $a(\lambda)$, $b(\lambda)$, $c_1(\lambda)$,
$d_1(\lambda)$ and $e_1(\lambda)$ are listed on table 2, and 
$f_1(\lambda)=1-b(\lambda)+d_1(\lambda)$. In ref. \cite{MA}  such vertex 
model was mapped on
a certain isotropic branch  of the Izergin-Korepin model, and consequently
its algebraic Bethe Ansatz solution is similar to that developed by Tarasov
\cite{TA} ( see also ref. \cite{MA} ). By including the inhomogeneities
$\{ \lambda_1, \cdots, \lambda_n \} $ we conclude that the 
corresponding eigenvalues are
\bear
&& \Lambda^{Osp(1|2)}\left(\lambda,\{\lambda_{i}\},\{\mu_{j}\}\right) =
\prod_{i=1}^{n} (\lambda-\lambda_{i}+1) \prod_{j=1}^{m} \frac{\mu_{j}-\lambda+1}{\mu_{j}-\lambda} +
\nonumber \\
&& \prod_{i=1}^{n} (\lambda-\lambda_{i}) \prod_{j=1}^{m}  -\frac{(\lambda-\mu_{j}-1/2)}{(\lambda-\mu_{j})} \frac{(\lambda-\mu_{j}+1)}{(\lambda-\mu_{j}+1/2)}+ 
\prod_{i=1}^{n} \frac{(\lambda-\lambda_{i})(\lambda-\lambda_{i}+1/2)}{(\lambda-\lambda_{i}+3/2)}
\prod_{j=1}^{m}  \frac{\lambda-\mu_{j}+3/2} {\lambda-\mu_{j}+1/2}
\nonumber \\
\ear
where the variables $\{\mu_{j}\}$ satisfy the equation
\EQ
\prod_{i=1}^{n} \frac{\mu_{j}-\lambda_{i}+1}{\mu_{j}-\lambda_{i}} = 
- \prod_{k=1}^{m} -\frac{(\mu_{j}-\mu_{k}-1/2)(\mu_{j}-\mu_{k}+1)}
{(\mu_{j}-\mu_{k}+1/2)(\mu_{j}-\mu_{k}-1)} , ~~ j=1, \dots,m.
\EN

Now we come to the point where all the results can be 
gathered together in the following way. Supposing we are 
interested in the eigenvalues of our original vertex model, we 
start with the eigenvalue formula 
(55) and use  recurrence relation (65) until we reach the 
problem of diagonalizing the 
$B_{1}$, $C_{1}$, $D_{2}$ and $Osp(1|2)$ models.  Then,
we have to take into 
account our results for the eigenvalues of these systems, which 
are collected in 
equations (68,71,75,78). By using this recipe,  it is straightforward to find
the eigenvalues expressions for the vertex 
models  we have so far discussed in this paper.  We now will
list our results for the eigenvalues.  
For the $B_n$ ($n \geq 1$) model we have
\bear
&& \Lambda^{B_n} \left(\lambda;\{\lambda^{(1)}_j\}, 
\cdots,\{\lambda^{(n)}_j\} \right) = 
\nonumber \\
&& [a(\lambda)]^{L}\prod_{j=1}^{m_{1}} \frac{\lambda-\lambda^{(1)}_{j}
-\frac{1}{2}} {\lambda-\lambda^{(1)}_{j}+\frac{1}{2}}+
[e_{n}(\lambda)]^{L}\prod_{j=1}^{m_{1}} \frac{\lambda-\lambda^{(1)}_{j}+n} 
{\lambda-\lambda^{(1)}_{j}+n-1}
 + [b(\lambda)]^{L}\sum_{l=1}^{2n-1} G_{l}(\lambda,\{ \lambda_j^{(\beta)} \})
\ear
where the functions $G_{l}(\lambda,\{ \lambda_j^{(\beta)} \})$ are given by 
\EQ
G_{l}(\lambda,\{ \lambda_j^{(\beta)} \} ) = \cases{
\displaystyle \prod_{j=1}^{m_l} \frac{\lambda-
\lambda^{(l)}_{j}+\frac{l+2}{2}} 
{\lambda-\lambda^{(l)}_{j}+\frac{l}{2}}
\prod_{k=1}^{m_{l+1}} \frac{\lambda-\lambda^{(l+1)}_{k}
+\frac{l-1}{2}}
{\lambda-\lambda^{(l+1)}_{k}+\frac{l+1}{2}},& $l=1, 
\cdots, n-1$ \cr
\displaystyle \prod_{k=1}^{m_n} \frac{(\lambda-\lambda^{(n)}_{k}+\frac{n-2}{2})}
{(\lambda-\lambda^{(n)}_{k}+\frac{n}{2})}
\frac{(\lambda-\lambda^{(n)}_{k}+\frac{n+1}{2})} 
{(\lambda-\lambda^{(n)}_{k}+\frac{n-1}{2})},& $l=n$ \cr
G_{2n-l}(1/2-n-\lambda,\{-\lambda_j^{(\beta)}\} ),&$ l =n+1,\cdots,2n-1$ \cr } 
\EN

For  $C_{n}$ ($n \geq 2$) we have
\bear
&& \Lambda^{C_n} \left(\lambda;\{\lambda^{(1)}_j \}, \cdots,\{\lambda^{(n)}_j\} \right) = \nonumber \\
&& [a(\lambda)]^L \prod_{j=1}^{m_1} \frac{\lambda-\lambda^{(1)}_j
-\frac{1}{2}}
{\lambda- \lambda^{(1)}_j+\frac{1}{2}} +
[e_n(\lambda)]^L \prod_{j=1}^{m_1} \frac{\lambda-\lambda^{(1)}_j+\frac{2n+3}{2}}
{\lambda- \lambda^{(1)}_j+\frac{2n+1}{2}} 
 + [b(\lambda)]^L \sum_{l=1}^{2n-2} G_{l}(\lambda,\{ \lambda_j^{(\beta)} \} )
\ear
where the functions $G_{l}(\lambda, \{ \lambda_j^{(\beta)} \} )$ are  
\EQ
G_{l}(\lambda,\{ \lambda_j^{(\beta)} \}) = \cases{ 
\displaystyle \prod_{j=1}^{m_{l}} 
\frac{\lambda-\lambda^{(l)}_{j}+\frac{l+2}{2}} 
{\lambda-\lambda^{(l)}_{j}+\frac{l}{2}}
\prod_{k=1}^{m_{l+1}} \frac{\lambda-\lambda^{(l+1)}_{k}+\frac{l-1}{2}}
{\lambda-\lambda^{(l+1)}_{k}+\frac{(l+1)}{2}},& $
l=1,\cdots,n-2 $ \cr
 \displaystyle \prod_{k=1}^{m_{n-1}} \frac{\lambda-\lambda^{(n-1)}_{k}+\frac{n+1}{2}}
{\lambda-\lambda^{(n-1)}_{k}+\frac{n-1}{2}}
\prod_{k=1}^{m_{n}}
\frac{\lambda-\lambda^{(n)}_{k}+\frac{n-3}{2}} 
{\lambda-\lambda^{(n)}_{k}+\frac{n+1}{2}}, & $ l=n-1$  \cr
G_{2n-1-l}(-1-n-\lambda,\{-\lambda_j^{(\beta)} \} ),& $ l=n, \cdots, 2n-2$ \cr }
\EN

For $D_{n} $($n \geq 3)$ we have
\bear
&& \Lambda^{D_n} \left(\lambda;\{\lambda^{(1)}_j \}, \cdots,\{\lambda^{(n)}_j\} \right) = \nonumber \\
&& [a(\lambda)]^L \prod_{j=1}^{m_1} \frac{\lambda-\lambda^{(1)}_j-\frac{1}{2}}
{\lambda- \lambda^{(1)}_j+\frac{1}{2}} +
[e_n(\lambda)]^L \prod_{j=1}^{m_1} \frac{\lambda-\lambda^{(1)}_j+\frac{2n-1}{2}}
{\lambda- \lambda^{(1)}_j+\frac{2n-3}{2}} 
 + [b(\lambda)]^L \sum_{l=1}^{2n-2} G_{l}(\lambda, \{ \lambda_j^{(\beta)} \})
\ear
where the functions $G_{l}(\lambda,\{ \lambda_j^{(\beta)} \})$ are  
\EQ
G_{l}(\lambda,\{\lambda_j^{(\beta)} \}) =  \cases{
\displaystyle \prod_{j=1}^{m_{l}} \frac{\lambda-\lambda^{(l)}_{j}
+\frac{l+2}{2}} 
{\lambda-\lambda^{(l)}_{j}+\frac{l}{2}}
\prod_{k=1}^{m_{l+1}} \frac{\lambda-\lambda^{(l+1)}_{k}+\frac{l-1}{2}}
{\lambda-\lambda^{(l+1)}_{k}+\frac{l+1}{2}},& $
l=1,\dots,n-3 $ \cr
\displaystyle \prod_{j=1}^{m_{n-2}} \frac{\lambda-\lambda^{(n-2)}_{j}+\frac{n}{2}}
{\lambda-\lambda^{(n-2)}_{j}+\frac{n-2}{2}}
\prod_{j=1}^{m_+}
\frac{\lambda-\lambda^{(+)}_{j}+\frac{n-3}{2}}
{\lambda-\lambda^{(+)}_{j}+\frac{n-1}{2}}
\prod_{j=1}^{m_{-}}
\frac{\lambda-\lambda^{(-)}_{j}+\frac{n-3}{2}}
{\lambda-\lambda^{(-)}_{j}+\frac{n-1}{2}},& $l=n-2$ \cr
\displaystyle \prod_{j=1}^{m_{+}}
\frac{\lambda-\lambda^{(+)}_{j}+\frac{n-3}{2}}
{\lambda-\lambda^{(+)}_{j}+\frac{n-1}{2}}
\prod_{j=1}^{m_{-}}
\frac{\lambda-\lambda^{(-)}_{j}+\frac{n+1}{2}}
{\lambda-\lambda^{(-)}_{j}+\frac{n-1}{2}},& $l=n-1$ \cr
G_{2n-1-l}(1-n-\lambda,\{-\lambda_j^{(\beta)} \}),& $l=n,\cdots,2n-2$ \cr }
\EN

For $Osp(2n-1|2)$ ($n \geq 2)$ we have
\bear
&& \Lambda^{Osp(2n-1|2)} \left(\lambda;\{\lambda^{(1)}_j \}, 
\cdots,\{\lambda^{(n)}_j\} \right) = 
\nonumber \\
&& [a(\lambda)]^L \prod_{j=1}^{m_1} -\frac{\lambda-\lambda^{(1)}_{j}+\frac{1}{2}}
{\lambda-\lambda^{(1)}_{j}-\frac{1}{2}} +
[e_n(\lambda)]^L \prod_{j=1}^{m_1} -\frac{\lambda-\lambda^{(1)}_{j}+n-3 }
{\lambda-\lambda^{(1)}_{j}+n-2} +
[b(\lambda)]^L \sum_{l=1}^{2n-1} G_{l}(\lambda,\{ \lambda_j^{(\beta)} \}  )
\nonumber \\
\ear
where the functions $G_{l}(\lambda,\{ \lambda_j^{(\beta)} \})$ are  
\EQ
G_{l}(\lambda,\{ \lambda_j^{(\beta)} \}) = \cases{
\displaystyle \prod_{j=1}^{m_{l}} \frac{\lambda-\lambda^{(l)}_{j}+\frac{l}{2}} 
{\lambda-\lambda^{(l)}_{j}+\frac{l-2}{2}}
\prod_{k=1}^{m_{l+1}} \frac{\lambda-\lambda^{(l+1)}_{k}+\frac{l-3}{2}}
{\lambda-\lambda^{(l+1)}_{k}+\frac{l-1}{2}},& $
l=1,\cdots,n-1 $ \cr
\displaystyle \prod_{k=1}^{m_{n}} \frac{(\lambda-\lambda^{(n)}_{k}+\frac{n-1}{2})}
{(\lambda-\lambda^{(n)}_{k}+\frac{n-3}{2})}
\frac{(\lambda-\lambda^{(n)}_{k}+\frac{n-4}{2})} 
{(\lambda-\lambda^{(n)}_{k}+\frac{n-2}{2})},& $l=n$ \cr
 G_{2n-l}(5/2-n-\lambda,\{-\lambda_j^{(\beta)} \}),& $ l=n+1, \cdots, 2n-1$ \cr }
\EN

For $Osp(2|2n-2)$, $(n \geq 2)$ we have
\bear
&& \Lambda^{Osp(2|2n-2)} \left(\lambda;\{\lambda^{(1)}_j \}, 
\cdots,\{\lambda^{(n)}_j\} \right) = 
\nonumber \\
&& [a(\lambda)]^L \prod_{j=1}^{m_1} -\frac{\lambda-\lambda^{(1)}_{j}+\frac{1}{2}}
{\lambda-\lambda^{(1)}_{j}-\frac{1}{2}} +
[e_n(\lambda)]^L \prod_{j=1}^{m_1} -\frac{\lambda-\lambda^{(1)}_{j}+\frac{2n-3}{2}}
{\lambda-\lambda^{(1)}_{j}+\frac{2n-1}{2}} +
[b(\lambda)]^L \sum_{l=1}^{2n-2} G_{l}(\lambda,\{\lambda_j^{(\beta)} \})
\nonumber \\
\ear
where the functions $G_{l}(\lambda,\{ \lambda_j^{(\beta)} \})$ are  
\EQ
G_{l}(\lambda,\{ \lambda_j^{(\beta)} \}) = \cases{
\displaystyle \prod_{j=1}^{m_{l}} \frac{\lambda-\lambda^{(l)}_{j}+\frac{l}{2}} 
{\lambda-\lambda^{(l)}_{j}+\frac{l-2}{2}}
\prod_{k=1}^{m_{l+1}} \frac{\lambda-\lambda^{(l+1)}_{k}+\frac{l-3}{2}}
{\lambda-\lambda^{(l+1)}_{k}+\frac{l-1}{2}},& $
l=1,\cdots,n-2 $ \cr
\displaystyle \prod_{j=1}^{m_{n-1}} \frac{\lambda-\lambda^{(n-1)}_{j}+\frac{n-1}{2}}
{\lambda-\lambda^{(n-1)}_{j}+\frac{n-3}{2}}
 \prod_{k=1}^{m_{n}}
\frac{\lambda-\lambda^{(n)}_{k}+\frac{n-5}{2}} 
{\lambda-\lambda^{(n)}_{k}+\frac{n-1}{2}},& $ l=n-1 $ \cr
G_{2n-1-l}(1-n-\lambda,\{-\lambda_j^{(\beta)} \} ),& $ l=n,\cdots,2n-2 $ \cr }
\EN

For $Osp(2n-2|2)$ ($n\geq 3)$ we have
\bear
&& \Lambda^{Osp(2n-2|2)} \left( \lambda;\{\lambda^{(1)}_j \}, 
\cdots,\{\lambda^{(n)}_j\} \right) = \nonumber \\
&& [a(\lambda)]^L \prod_{j=1}^{m_1} -\frac{\lambda-\lambda^{(1)}_j+\frac{1}{2}}
{\lambda- \lambda^{(1)}_j-\frac{1}{2}} +
[e_n(\lambda)]^L \prod_{j=1}^{m_1} -\frac{\lambda-\lambda^{(1)}_j+\frac{2n-7}{2}}
{\lambda- \lambda^{(1)}_j+\frac{2n-5}{2}} 
 + [b(\lambda)]^L \sum_{l=1}^{2n-2} G_{l}(\lambda,\{ \lambda_j^{(\beta)} \})
\nonumber \\
\ear
where the functions $G_{l}(\lambda,\{ \lambda_j^{(\beta)} \})$ are 
\EQ
G_{l}(\lambda,\{ \lambda_j^{(\beta)} \}) = \cases{
\displaystyle \prod_{j=1}^{m_{l}} \frac{\lambda-\lambda^{(l)}_{j}+\frac{l}{2}} 
{\lambda-\lambda^{(l)}_{j}+\frac{l-2}{2}}
\prod_{k=1}^{m_{l+1}} \frac{\lambda-\lambda^{(l+1)}_{k}+\frac{l-3}{2}}
{\lambda-\lambda^{(l+1)}_{k}+\frac{l-1}{2}},& $
l=1,\dots,n-3 $ \cr
 \displaystyle \prod_{j=1}^{m_{n-2}} \frac{\lambda-\lambda^{(n-2)}_{j}+\frac{n-2}{2}}
{\lambda-\lambda^{(n-2)}_{j}+\frac{n-4}{2}}
\prod_{j=1}^{m_{+}}
\frac{\lambda-\lambda^{(+)}_{j}+\frac{n-5}{2}}
{\lambda-\lambda^{(+)}_{j}+\frac{n-3}{2}}
\prod_{j=1}^{m_{-}}
\frac{\lambda-\lambda^{(-)}_{j}+\frac{n-5}{2}}
{\lambda-\lambda^{(-)}_{j}+\frac{n-3}{2}},& $ l=n-2$ \cr
\displaystyle \prod_{j=1}^{m_{+}}
\frac{\lambda-\lambda^{(+)}_{j}+\frac{n-5}{2}}
{\lambda-\lambda^{(+)}_{j}+\frac{n-3}{2}}
\prod_{j=1}^{m_{-}}
\frac{\lambda-\lambda^{(-)}_{j}+\frac{n-1}{2}}
{\lambda-\lambda^{(-)}_{j}+\frac{n-3}{2}},& $l=n-1 $ \cr
G_{2n-1-l}(3-n-\lambda,\{ -\lambda_j^{(\beta)} \} ),& $ l=n,\cdots,2n-2$ \cr }
\EN

Finally, for  $Osp(1|2n)$ ($n \geq 1)$ we have
\bear
&& \Lambda^{Osp(1|2n)} \left(\lambda;\{\lambda^{(1)}_j \}, 
\cdots,\{\lambda^{(n)}_j\} \right) = 
\nonumber \\
&& [a(\lambda)]^L \prod_{j=1}^{m_1} \frac{\lambda-\lambda^{(1)}_{j}-\frac{1}{2}}
{\lambda-\lambda^{(1)}_{j}+\frac{1}{2}} +
[e_n(\lambda)]^L \prod_{j=1}^{m_1} \frac{\lambda-\lambda^{(1)}_{j}+n+1}
{\lambda-\lambda^{(1)}_{j}+n} +
[b(\lambda)]^L \sum_{l=1}^{2n-1} G_{l}(\lambda,\{ \lambda_j^{(\beta)} \})
\ear
where the functions $G_{l}(\lambda,\{ \lambda_j^{(\beta)} \})$  are
\EQ
G_{l}(\lambda,\{ \lambda_j^{(\beta)} \}) = \cases{
\displaystyle \prod_{j=1}^{m_{l}} \frac{\lambda-\lambda^{(l)}_{j}+\frac{l+2}{2}} 
{\lambda-\lambda^{(l)}_{j}+\frac{l}{2}}
\prod_{k=1}^{m_{l+1}} \frac{\lambda-\lambda^{(l+1)}_{k}+\frac{l-1}{2}}
{\lambda-\lambda^{(l+1)}_{k}+\frac{l+1}{2}},& $
l=1,\cdots,n-1 $ \cr
\displaystyle \prod_{k=1}^{m_{n}} -\frac{(\lambda-\lambda^{(n)}_{k}+\frac{n-1}{2})}
{(\lambda-\lambda^{(n)}_{k}+\frac{n+1}{2})}
\frac{(\lambda-\lambda^{(n)}_{k}+\frac{n+2}{2})} 
{(\lambda-\lambda^{(n)}_{k}+\frac{n}{2})},& $ l=n$ \cr
G_{2n-l}(-1/2-n-\lambda,\{ -\lambda_j^{(\beta)} \} ),& $ l=n+1, \cdots, 2n-1$ \cr}
\EN

We see that the eigenvalues depend on the parameters 
$\left\{ \{\lambda^{(1)}_{j}\},\{\lambda^{(2)}_{j}\}, \cdots, \{\lambda^{(n)}_{j}\} \right\}$, which   
represent the multi-particle Hilbert space  of the many 
steps necessary for the diagonalization of the nesting 
problem. In the  expressions (80-93) we have 
already performed convenient shifts in these parameters, 
$\{ \lambda_j^{(\beta)} \}
\rightarrow \{ \lambda_j^{(\beta)} \} - \delta^{(\beta)}$, 
in order to present the final results
in a more symmetrical way. In table 4 we show the values for the shifts
$\delta^{(\beta)} $.
These set of variables are constrained by the Bethe Ansatz equation,
again at each level of the nesting. The same 
recipe described above for the eigenvalues also works for determining the corresponding 
Bethe Ansatz equations. We start with equation (57) and 
each step of the nesting is 
disentangled by using the recurrence relation (66). When  we reach
the last step, we take into account the Bethe Ansatz results (69,72,76,79) 
for the inhomogeneous  
$B_{1}$, $C_{1}$, $D_{2}$ and $Osp(1|2)$ vertex models. It turns out that the
nested 
Bethe Ansatz equations have the following structure. All the vertex models share 
a common part, which can be written as
\EQ
\prod_{k=1}^{m_{l-1}} \frac{\lambda^{(l)}_{j}-\lambda^{(l-1)}_{k}
+1/2}{\lambda^{(l)}_{j}-\lambda^{(l-1)}_{k}-1/2} 
\prod_{k=1,\;k \neq j}^{m_{l}} \frac{\lambda^{(l)}_{j}-\lambda^{(l)}_{k}-1}
{\lambda^{(l)}_{j}-\lambda^{(l)}_{k}+1}
\prod_{k=1}^{m_{l+1}} \frac{\lambda^{(l)}_{j}-\lambda^{(l+1)}_{k}+1/2}
{\lambda^{(l)}_{j}-\lambda^{(l+1)}_{k}-1/2} = 1, ~~ l= 2,\cdots, s(n)
\EN
where $s(n)= n-1$ for the $B_{n}$, $Osp(2n-1|2)$ and $Osp(1|2n)$ models 
; $s(n)=n-2$ for the $C_{n}$,  
$Osp(2|2n-2)$ models; $s(n)=n-3$ for the $D_n$ and $Osp(2n-2|2)$ models. 
The remaining equations are somewhat model dependent and below we list their 
particular forms. For the models $B_{n}$ $(n \geq 2)$, $C_{n}$ $(n \geq 3)$, 
$D_{n}$ $(n \geq 4)$ and $Osp(1|2n)$ $(n \geq 2)$ the equation for the first root $\{\lambda^{(1)}_{j}\}$ is given by
\EQ
\left[ \frac{\lambda^{(1)}_{j}-1/2}{\lambda^{(1)}_{j}+1/2} \right]^{L} = 
\prod_{k=1,\; k \neq j}^{m_{1}} \frac{\lambda^{(1)}_{j}-\lambda^{(1)}_{k}-1}{\lambda^{(1)}_{j}-\lambda^{(1)}_{k}+1}
\prod_{k=1}^{m_{2}} \frac{\lambda^{(1)}_{j}-\lambda^{(2)}_{k}+1/2}{\lambda^{(1)}_{j}-\lambda^{(2)}_{k}-1/2}
\EN
while for the $Osp(2n-1|2)$ ($n \geq 2)$ , 
$Osp(2|2n-2)$ ($n \geq 3)$ and $Osp(2n-2|2)$ ($ n \geq 4$) we have
\EQ
\left[ \frac{\lambda^{(1)}_{l}+1/2}{\lambda^{(1)}_{l}-1/2} \right]^{L}=
(-1)^{L-m_{1}-1}
\prod_{k=1}^{m_{2}} \frac{\lambda^{(1)}_{l}-\lambda^{(2)}_{k}+1/2}{\lambda^{(1)}_{l}-\lambda^{(2)}_{k}-1/2} 
\EN

Due to the peculiar root structure of the models $C_2$, $D_3$  and $Osp(2|2)$,
$Osp(4|2)$, their
first Bethe Ansatz equations  
are a bit different than that presented in equations (95) and (96),
respectively. In order to avoid further confusion we have to   
quote  them separately. For the $C_2$ model the Bethe Ansatz equation for
$\{ \lambda_j^{(1)}\} $ is
\EQ
\left[ \frac{\lambda^{(1)}_{j}-1/2}{\lambda^{(1)}_{j}+1/2} \right]^{L} = 
\prod_{k=1,\; k \neq j}^{m_{1}} \frac{\lambda^{(1)}_{j}-\lambda^{(1)}_{k}-1}{\lambda^{(1)}_{j}-\lambda^{(1)}_{k}+1}
\prod_{k=1}^{m_{2}} \frac{\lambda^{(1)}_{j}-\lambda^{(2)}_{k}+1}{\lambda^{(1)}_{j}-\lambda^{(2)}_{k}-1} 
\EN
while for the $Osp(2|2)$ model we have
\EQ
\left[ \frac{\lambda^{(1)}_{l}+1/2}{\lambda^{(1)}_{l}-1/2} \right]^{L}
 = (-1)^{L-m_{1}-1}
\prod_{k=1}^{m_{2}} \frac{\lambda^{(1)}_{l}-\lambda^{(2)}_{k}+1}{\lambda^{(1)}_{l}-\lambda^{(2)}_{k}-1} 
\EN

Furthermore, the first equation for the $D_3$ model is
\EQ
\left[ \frac{\lambda^{(1)}_{j}-1/2}{\lambda^{(1)}_{j}+1/2} \right]^{L} = 
\prod_{k=1,\; k \neq j}^{m_{1}} \frac{\lambda^{(1)}_{j}-\lambda^{(1)}_{k}-1}{\lambda^{(1)}_{j}-\lambda^{(1)}_{k}+1}
\prod_{k=1}^{m_{+}} \frac{\lambda^{(1)}_{j}-\lambda^{(+)}_{k}+1/2}{\lambda^{(1)}_{j}-\lambda^{(+)}_{k}-1/2} 
\prod_{k=1}^{m_{-}} \frac{\lambda^{(1)}_{j}-\lambda^{(-)}_{k}+1/2}{\lambda^{(1)}_{j}-\lambda^{(-)}_{k}-1/2} 
\EN
and for the model $Osp(4|2)$  we have
\EQ
\left[ \frac{\lambda^{(1)}_{l}+1/2}{\lambda^{(1)}_{l}-1/2} \right]^{L}
 = (-1)^{L-m_{1}-1}
\prod_{k=1}^{m_{+}} \frac{\lambda^{(1)}_{l}-\lambda^{(+)}_{k}+1/2}{\lambda^{(1)}_{l}-\lambda^{(+)}_{k}-1/2}
\prod_{k=1}^{m_{-}} \frac{\lambda^{(1)}_{l}-\lambda^{(-)}_{k}+1/2}{\lambda^{(1)}_{l}-\lambda^{(-)}_{k}-1/2}
\EN

The Bethe Ansatz equation 
for the last variables are common for the pairs
 $\{B_{n}, Osp(2n-1|2) \}$, 
$\{C_{n}, Osp(2|2n-2)\}$, $\{D_{n}, Osp(2n-2|2)\}$ and are given as follows.
For $B_{n}$ and $Osp(2n-1|2)$ models the variables
$\{\lambda^{(n)}_{j}\}$ satisfy the equation
\EQ
\prod_{k=1}^{m_{n-1}} \frac{\lambda^{(n)}_{j}-\lambda^{(n-1)}_{k}+1/2}{\lambda^{(n)}_{j}-\lambda^{(n-1)}_{k}-1/2} 
\prod_{k=1,\; k \neq j}^{m_{n}} \frac{\lambda^{(n)}_{j}-\lambda^{(n)}_{k}-1/2}{\lambda^{(n)}_{j}-\lambda^{(n)}_{k}+1/2} = 1
\EN

For $C_{n}$ and $Osp(2|2n-2)$, the parameters 
$\{\lambda^{(n-1)}_{j}\}$ and $\{\lambda^{(n)}_{j}\}$ satisfy the equations 
\footnote{ For $n=2$ ( $C_2$ and $Osp(2|2)$ models ) we have to consider only the last equation given in (102) .}
\bear
\prod_{k=1}^{m_{n-2}} \frac{\lambda^{(n-1)}_{j}-\lambda^{(n-2)}_{k}+1/2}{\lambda^{(n-1)}_{j}-\lambda^{(n-2)}_{k}-1/2} 
\prod_{k=1,\;  k \neq j }^{m_{n-1}} \frac{\lambda^{(n-1)}_{j}-\lambda^{(n-1)}_{k}-1}{\lambda^{(n-1)}_{j}-\lambda^{(n-1)}_{k}+1}
\prod_{k=1}^{m_{n}} \frac{\lambda^{(n-1)}_{j}-\lambda^{(n)}_{k}+1}{\lambda^{(n-1)}_{j}-\lambda^{(n)}_{k}-1} = 1, \nonumber \\
\prod_{j=1}^{m_{n-1}} \frac{\lambda^{(n)}_{l}-\lambda^{(n-1)}_{j}-1}{\lambda^{(n)}_{l}-\lambda^{(n-1)}_{j}+1} 
\prod_{k=1,\; k \neq j}^{m_{n}} \frac{\lambda^{(n)}_{j}-\lambda^{(n)}_{k}+2}{\lambda^{(n)}_{j}-\lambda^{(n)}_{k}-2} = 1.
\ear

For $D_{n}$ and $Osp(2n-2|2)$, the parameters 
$\{\lambda^{(n-2)}_{j}\}$, 
$\{\lambda^{(+)}_{j}\}$ and $\{\lambda^{(-)}_{j}\}$ satisfy  the equations\footnote{Here we  have assumed the identifications 
$\lambda_j^{(n-1)} \equiv \lambda_j^{(+)} $ and
$\lambda_j^{(n)} \equiv \lambda_j^{(-)} $. Moreover, for $n=3$ ($D_3$ and $Osp(4|2)$ models ) only the last equation in (103) matters. } 
\bear
\prod_{\gamma=\pm} \prod_{k=1}^{m_{\gamma}} 
\frac{\lambda^{(n-2)}_{j}-\lambda^{(\gamma)}_{k}+1/2}{\lambda^{(n-2)}_{j}-\lambda^{(\gamma)}_{k}-1/2} 
\prod_{k=1,\; k \neq j}^{m_{n-2}} \frac{\lambda^{(n-2)}_{j}-\lambda^{(n-2)}_{k}-1}{\lambda^{(n-2)}_{j}-\lambda^{(n-2)}_{k}+1}
\prod_{k=1}^{m_{n-3}} \frac{\lambda^{(n-2)}_{j}-\lambda^{(n-3)}_{k}+1/2}{\lambda^{(n-2)}_{j}-\lambda^{(n-3)}_{k}-1/2} = 1, \nonumber \\
\prod_{k=1}^{m_{n-2}} \frac{\lambda^{(\pm)}_{j}-\lambda^{(n-2)}_{k}+1/2}{\lambda^{(\pm)}_{j}-\lambda^{(n-2)}_{k}-1/2} 
\prod_{k=1,\; k \neq j}^{m_{\pm}} \frac{\lambda^{(\pm)}_{j}-\lambda^{(\pm)}_{k}-1}{\lambda^{(\pm)}_{j}-\lambda^{(\pm)}_{k}+1} = 1
\ear
and finally for the $Osp(1|2n)$ vertex model  
the equation for variables $\{\lambda^{(n)}_j \}$ is\footnote{We remark that
the Bethe Ansatz equations and eigenvalues for the models $B_1$, $C_1$, $D_2$
and $Osp(1|2)$ can be obtained from equations (68-69,71-72,75-76,78-79) by setting the inhomogeneities to zero.}  
\EQ
\prod_{k=1}^{m_{n-1}} \frac{\lambda^{(n)}_{j}-\lambda^{(n-1)}_{k}+1/2}{\lambda^{(n)}_{j}-\lambda^{(n-1)}_{k}-1/2} 
\prod_{k=1}^{m_{n}} -\frac{(\lambda^{(n)}_{j}-\lambda^{(n)}_{k}+1/2)(\lambda^{(n)}_{j}-\lambda^{(n)}_{k}-1)}
{(\lambda^{(n)}_{j}-\lambda^{(n)}_{k}-1/2)(\lambda^{(n)}_{j}-\lambda^{(n)}_{k}+1)} = 1
\EN

We would like  to close this section with the following remarks. We begin by
discussing and comparing our results to previous work in the literature. It is
not difficult to check that our results for the Bethe Ansatz equations are
equivalent to the analyticity of the eigenvalues as a function of variables
$\left \{ \lambda_{j}^{(1)} \}, \cdots, \{\lambda_{j}^{(n)} \} \right \} $,
i.e. all the residues on the direct and crossed poles vanish. This is 
precisely the main important ingredient entering in the framework of the
analytical Bethe Ansatz. Indeed, the results of this paper for the eigenvalues
and the Bethe equations of the $B_n$, $C_n$, $D_n$ and $Osp(1|2n)$ \footnote{
We remark that ref.\cite{MA} have used the grading $f \cdots fbf \cdots f$ and here
we have taken the sequence $b \cdots bfb \cdots b$. This is the reason why
the phase factors of ref.\cite{MA} and that of equation (104) are different. We
also have noticed misprints in ref.\cite{RE}  concerning some 
eigenvalue expressions.} models are in accordance to those obtained from
the analytical Bethe Ansatz approach in  
refs.\cite{RE,RE1,SU,MA}. As we have already  
commented in the introduction, the $D_n$ vertex model has also been solved 
by the algebraic Bethe Ansatz in ref.\cite{KV}.  These authors argued 
that the nesting problem, in a convenient basis, could be transformed to two 
commuting eigenvalue problems. One of them has the permutation operator as
the main intertwiner, and the algebraic Bethe solution goes along to that
known to work for the multi-state $6$-vertex models \cite{KS,BA,KRA}. The other one was
related to the Temperley-Lieb operator ( in our notation of section 2 ), 
no explicit algebraic solution was attempted, and
the eigenvalue results were obtained via the crossing property.
In our approach, however, we deal with
these two operators together in the diagonalization problem and the explicit
use of crossing it is not needed.  We also noticed that while 
our formulation works
for vertex models having both even or odd numbers of states $q$ per link, the
basis used in ref.\cite{KV} appears to be suitable only for $q$ even. Since
our approach and the one of ref. \cite{KV} are quite different from the
very beginning, we were not able to find a simple way of comparing the
results for the eigenvectors.

For the models based on the superalgebras $Osp(n|2m)$, we notice that 
the results for the eigenvalues and Bethe Ansatz equations present some
additional phase factors when compared to that obtained for the pure
``bosonic'' $B_n$, $C_n$ and $D_n$ models. These phases are the only
liquid difference between  the standard and the graded formulations of the
quantum inverse scattering method. They are indeed twisted boundary conditions
(periodic and antiperiodic) having a fermionic sector dependence ( see e.g.
ref. \cite{MA} ). In appendix $C$ we show how such phase factors can be
absorbed in the supersymmetric formulation of the $Osp(2n-1|2)$,$Osp(2|2n-2)$,
$Osp(2n-2|2)$ and $Osp(1|2n)$ vertex models. Without these extra factors, the
nested Bethe Ansatz equations becomes more symmetrical, similar to what 
happens for the ``bosonic'' models. It turns out that such symmetrical
Bethe Ansatz equations can even be formulated in a more compact form, 
reflecting the underlying group symmetry of these vertex models. In order to
see this, we first need to make the following definitions. Let 
$C_{ab}$ be 
Cartan matrix associated to the Dynkin diagrams of figure 1, 
and $\eta_{a}$ 
the normalized length of the root of type $\beta_a$. Here we assume
that the length of the long root is $2$. This means that
$\eta_{a} = 
\frac{2}{(\alpha_{a},\alpha_{a})}$ and $\eta_a=1$ 
for a long root \footnote{More precisely, for 
$D_{n}$ and $Osp(2n-2|2)$ $\eta_{a}=1$, $a=1, \dots, n$; for $B_{n}$ and $Osp(2n-1|2)$  
$\eta_{a}=1$, $a=1, \dots, n-1$, and $\eta_{n}=2$; 
for $C_{n}$ and $Osp(2|2n-2)$  
$\eta_{a}=2$, $a=1, \dots, n-1$, and $\eta_{n}=1$ (see for instance
ref.\cite{CO}). }. Taking
into account these definitions, it is possible to rewrite the Bethe
Ansatz equations (94-103)  
as\footnote{We notice that for the $C_{n}$ and $Osp(2|2n-2)$ vertex models one should rescale all $\{\lambda^{(a)}_{j}\}$ by a factor 2, $\{\lambda^{(a)}_{j}\} \rightarrow \frac{\{\lambda^{(a)}_{j}\}}{2}$ in equation (105) to recover
the previous results (94,95,97,98,100).}
\EQ
\left[
\frac{\lambda^{(a)}_{j} -\frac{\delta_{a,1}}{2\eta_{a}}}{\lambda^{(a)}_{j} +\frac{\delta_{a,1}}{2\eta_{a}}} 
\right]^{L} =
\prod_{b=1}^{r(n)} \prod_{k=1,\; k \neq j}^{m_{b}}
\frac{\lambda^{(a)}_{j}-\lambda^{(b)}_{k} -\frac{C_{a,b}}{2\eta_{a}}}{\lambda^{(a)}_{j}-\lambda^{(b)}_{k} +\frac{C_{a,b}}{2\eta_{a}}}, ~~ j=1, \dots, m_{a} ;~~ a=1, \dots, r(n)
\EN
where in general $r(n)$ is the number of roots of the underlying algebra. For
the vertex models solved in this papers $r(n) =n$.

The idea that the Bethe Ansatz equations can be 
compactly written in terms of their 
corresponding Lie algebra goes back to 
the work of Ogievestsky, Reshetikhin and Wiegmann \cite{OWR} who 
have conjectured similar formulas for the Bethe Ansatz equations 
of factorizable $S$-matrices based on the 
standard Lie algebras $A_{n}$, $B_{n}$, $C_{n}$, 
$D_{n}$, $E_{6}$, $E_{7}$, $E_{8}$, $F_{4}$, $G_{2}$ . 
Thus, our results can be seen as a rigorous proof of 
this conjecture for the rational $B_{n}$, $C_{n}$, 
$D_{n}$ vertex models. Moreover, together with the 
results of 
ref. \cite{KU} for the $Sl(n|m)$ algebra (see also 
ref.\cite{KE}), they also show that  this
conjecture can be extended to the case of superalgebras. 
An exception in this construction is again the $Osp(1|2n)$ vertex model. In 
order to  fit equation (105), we need to give a special meaning to the
black bullet $\bullet$ of figure 1 ( root $ \{ \lambda_j^{(n)} \} $ in the
Bethe Ansatz ). One could interpret it as a peculiar two-body self interaction
for root $\{ \lambda_j^{(n)} \} $, as the one present in the right hand side 
of equation (104).

	Our last remark is concerned with the underlying spin chain associated 
to the braid-monoid vertex models. The $R$-matrix 
(4) is regular at $\lambda=0$, 
thus the local conserved charges can be obtained through the 
logarithmic expansion of the corresponding transfer-matrix 
around $\lambda=0$. 
The first charge is the momentum itself  (we are assuming periodic boundary 
conditions) and the next one, i.e. the logarithmic derivative of $ T
(\lambda)$ at $\lambda=0$, is the Hamiltonian. From the expression of the 
$R$-matrix $R(\lambda)$ we found
\EQ
{\cal H} = \sum_{i=1}^{L} P^{g}_{i}+\frac{1}{\Delta} E_{i}
\EN

Analogously, the eigenenergies $E(L)$ of the corresponding Hamiltonian (106) 
can be calculated by taking the logarithmic derivative of 
$\Lambda(\lambda,\{\lambda^{(1)}_{i}\}, \dots ,\{\lambda^{(n)}_{j}\})$ 
at the regular point $\lambda=0$. The eigenenergies 
$E(L)$ are parametrized in terms of the variables 
$\{\lambda^{(1)}_{i}\}$ by\footnote{ It has become a tradition in the literature
to normalize the Bethe Ansatz variables by a pure imaginary factor as
$\lambda_j^{(\beta)} \rightarrow 
\frac{\lambda_j^{(\beta)}}{i} $. In this case, equation (107) reads
$ E(L) = \displaystyle \mp \sum_{i=1}^{m_{1}} \frac{1}{[\lambda^{(1)}_{i}]^{2} + 1/4} \pm L$.}
\EQ
E(L) = \pm \sum_{i=1}^{m_{1}} \frac{1}{[\lambda^{(1)}_{i}]^{2} - 1/4} \pm L
\EN
where the signs $\pm$ is related to the two possibilities for the 
Boltzmann weights of type $a(\lambda)$, i.e., $a(\lambda)=1 \pm \lambda$. The 
variables  $\{\lambda^{(1)}_{i}\}$ satisfy the general nested Bethe Ansatz 
equations (94-104), and therefore the spectrum of all $Osp(n|2m)$ chains can be 
determined solving these equations. A preliminary study of the 
root structures of the Bethe Ansatz equations (94-104) have shown an intricate 
behaviour for the general $Osp(n|2m)$. We leave a detailed analysis of these 
root structure, as well the ground state and the low-lying excitations for a 
forthcoming paper.
\section{Conclusions}

In this paper we have solved 
exactly a series of rational vertex models based on 
the braid-monoid algebra from a rather unified perspective.  The
general construction for the $R$-matrices presented in section 2
played an important role in many steps of our formulation of
the quantum inverse scattering method. In particular, it is notable how
such construction becomes useful to determine the universal character of
the fundamental commutations rules presented in section 3. It appears natural
to us to blame the properties of the braid-monoid algebra as 
the main mathematical structure behind such general picture. 

The commutation relations allowed us
to determine the
eigenvectors and the eigenvalues of the transfer matrix of the vertex models invariant by 
the  $B_{n}$, $C_{n}$, $D_{n}$, 
$Osp(2n-1|2)$, $Osp(2|2n-2)$, $Osp(2n-2|2)$ and $Osp(1|2n)$ symmetries from
systematic point of view.  As a consequence, we have been able to 
establish the 
solution of the general $Osp(n|2m)$ invariant spin chain. Their Bethe
Ansatz equations have been formulated in terms of basic properties of
the underlying group symmetry such as the root structure. This means that,
in principle,
by solving equations (105) together with (107) we can determine
the spectrum of the general $Osp(n|2m)$ spin chain (106).

One possible extension of this work is to consider the 
trigonometric analogs of the vertex models solved in this paper. In general, we expect
 that the trigonometric vertex models based on the Birman-Wenzl-Murakami algebra\cite{WA,BWK} 
can be solved by introducing few adaptations to our construction of sections
3 and 4. 
In addition, 
we have noticed that similar ideas also works for 
the one-dimensional Hubbard model .
In this case, however, some other peculiarities such as the non-additive property of the $R$-matrix 
need to be disentangled, too. Due to the  special character of the Hubbard
$R$-matrix and its recognized importance 
in condensed matter 
physics, we  have 
dedicated a brief account of our results 
in a separated publication \cite{PMH}. It remains to be seen whether these
are isolated examples or our construction could even be more widely applicable
that one would think at first sight. The last possibility would suggest that
a more deep and essential mathematical structure is still to be grasped from
the Yang-Baxter algebra.

\section*{Acknowledgements}
The authors thank J. de Gier for helpful comments. 
This  
work was supported by  FOM (Fundamental Onderzoek der Materie) 
and Fapesp ( Funda\c c\~ao
de Amparo \'a Pesquisa do Estado de S. Paulo).

\centerline{\bf Appendix A : The two-particle state }
\setcounter{equation}{0}
\renewcommand{\theequation}{A.\arabic{equation}}

In this appendix we present some details  concerning the complete solution
of the eigenvalue problem for the two-particle state. Here we shall need
few extra commutation relations between the fields $\vec{B}(\lambda)$,
$\vec{B}^{*}(\lambda)$, $\vec{C}(\lambda)$ and $\vec{C^{*}}(\lambda)$. They
are given by
\EQ
C_{a}(\lambda)B_{b}(\mu) = 
B_{b}(\mu)C_{a}(\lambda) - \frac{1}
{b(\lambda-\mu)}[B(\lambda)A_{ab}(\mu) - B(\mu)A_{ab}(\lambda)]
\EN
\EQ
B^{*}_{a}(\lambda)B_{b}(\mu) =
B_{b}(\mu)B^{*}_{a}(\lambda) - 
\frac{1}{b(\lambda-\mu)}[F(\lambda)A_{ab}(\mu)  
-F(\mu)A_{ab}(\lambda)]
\EN
\EQ
C^{*}_{a}(\lambda)B_{b}(\mu) \ket{0} = 
\xi_{ab} \frac{d_{n}(\lambda-\mu)}{e_{n}(\lambda-\mu)}
\left[ B(\mu)D(\lambda) - A_{aa}(\lambda)A_{bb}(\mu) \right]  \ket{0}
\EN
where $a,b=1,\cdots,q-2$.

We start our discussion by considering the wanted terms for the two-particle
state. Since the terms proportional to 
$B_i(\lambda_1^{(1)}) B_j(\lambda_2^{(2)}) {\cal{F}}^{ji} $
have been already cataloged in section 4 ( see expression (55) ) we turn our
attention to those proportional to 
$-\frac{d_n(\lambda_1^{(1)}-\lambda_2^{(1)})}
{e_n(\lambda_1^{(1)}-\lambda_2^{(1)})}
F(\lambda_1^{(1)}) B(\lambda_2^{(1)}) 
\vec{\xi}. \vec{\cal{F}}$ . In other words, we have to show that the wanted 
terms proportional to the second part of the two-particle eigenvector (40) 
produce the eigenvalue expression (55) for $m_1=2$. It is not difficult to
collect the contributions proportional to $[a(\lambda)]^{L}$ and 
$[e_n(\lambda)^{L}]$. They are given respectively by
\EQ
\left[ \frac{a(\lambda^{(1)}_{1}-\lambda)}{e_{n}(\lambda^{(1)}_{1}-\lambda)} + 
\frac{1}{b(\lambda_2^{(1)}-\lambda)}
\frac{a(\lambda^{(1)}_{1}-\lambda)}{b(\lambda^{(1)}_{1}-\lambda)} 
\frac{d_{n}(\lambda^{(1)}_{1}-\lambda)}{e_{n}(\lambda^{(1)}_{1}-\lambda)}
\frac{e_{n}(\lambda^{(1)}_{1}-\lambda^{(1)}_{2})}{d_{n}(\lambda^{(1)}_{1}-\lambda^{(1)}_{2})} \right] 
[a(\lambda)]^{L} 
\EN
\EQ
\left[ \frac{a(\lambda-\lambda^{(1)}_{1})}{e_{n}(\lambda-\lambda^{(1)}_{1})} - 
\frac{1}{e_{n}(\lambda-\lambda^{(1)}_{1})} 
\frac{d_{n}(\lambda-\lambda^{(1)}_{2})}{e_{n}(\lambda-\lambda^{(1)}_{2})}
\frac{e_{n}(\lambda^{(1)}_{1}-\lambda^{(1)}_{2})}{d_{n}(\lambda^{(1)}_{1}-\lambda^{(1)}_{2})} \right] 
[e_{n}(\lambda)]^{L} 
\EN

Now if we use that the Boltzmann weights satisfy the following identities

\EQ
\left[ \frac{a(\lambda^{(1)}_{1}-\lambda)}{e_{n}(\lambda^{(1)}_{1}-\lambda)} + 
\frac{1}{b(\lambda^{(1)}_{2}-\lambda)} 
\frac{a(\lambda^{(1)}_{1}-\lambda)}{b(\lambda^{(1)}_{1}-\lambda)} 
\frac{d_{n}(\lambda^{(1)}_{1}-\lambda)}{e_n(\lambda^{(1)}_{1}-\lambda)}
\frac{e_{n}(\lambda^{(1)}_{1}-\lambda^{(1)}_{2})}{d_{n}(\lambda^{(1)}_{1}-\lambda^{(1)}_{2})} \right] = 
\prod_{i=1}^{2} \frac{a(\lambda^{(1)}_{i}-\lambda)}{b(\lambda^{(1)}_{i}-\lambda)}
\EN
\EQ
\left[ \frac{a(\lambda-\lambda^{(1)}_{1})}{e_{n}(\lambda-\lambda^{(1)}_{1})} - 
\frac{1}{e_{n}(\lambda-\lambda^{(1)}_{1})} 
\frac{d_{n}(\lambda-\lambda^{(1)}_{2})}{e_{n}(\lambda-\lambda^{(1)}_{2})}
\frac{e_{n}(\lambda^{(1)}_{1}-\lambda^{(1)}_{2})}{d_{n}(\lambda^{(1)}_{1}-\lambda^{(1)}_{2})} \right] =
\prod_{i=1}^{2} \frac{b(\lambda-\lambda^{(1)}_{i})}{e_{n}(\lambda-\lambda^{(1)}_{i})}
\EN
it is immediately seen that these two contributions are precisely the first
and the second terms of equation (55) for $m_1=2$.

The analysis of the wanted term proportional to $[b(\lambda)]^{L}$ is a bit
more involved. The basic reason for that is the presence of some ``asymmetric''
terms of kind 
$[b(\lambda) a(\lambda^{(1)}_{2})]^{L} F(\lambda^{(1)}_{1}) \sum_{ij}
\xi_{ij} {\cal{F}}^{ij}
$
and the peculiar form of the nested eigenvalue 
$\Lambda^{(1)}(\lambda,\{\lambda^{(1)}_{i}\})$.  We find that the terms 
contributing to this last wanted piece are given by
\EQ
-(q-2) \left[ 1 - \frac{1}{[b(\lambda-\lambda^{(1)}_{1})]^{2}} \right] \times
[b(\lambda)a(\lambda^{(1)}_{2})]^{L} 
\frac{d_{n}(\lambda^{(1)}_{1}-\lambda^{(1)}_{2})}{e_{n}(\lambda^{(1)}_{1}-\lambda^{(1)}_{2})}
F(\lambda^{(1)}_{1}) \sum_{ij} \xi_{ij} {\cal F}^{ji}
\EN
\EQ
-[b(\lambda) a(\lambda^{(1)}_{2})]^{L} 
\frac{1}{[b(\lambda-\lambda^{(1)}_{1})]^{2}}
\frac{d_{n}(\lambda-\lambda^{(1)}_{2})}{e_{n}(\lambda-\lambda^{(1)}_{2})}
\sum_{ijlm} \xi_{mj} [X^{(1)}(\lambda-\lambda^{(1)}_{1})]^{lm}_{li} F(\lambda^{(1)}_{1}) {\cal F}^{ji},
\EN
\EQ
-[b(\lambda) a(\lambda^{(1)}_{2})]^{L} 
\frac{d_{n}(\lambda-\lambda^{(1)}_{1})}{e_{n}(\lambda-\lambda^{(1)}_{1})}
\frac{[1-\hat{t}b(\lambda-\lambda^{(1)}_{1})]}
{b(\lambda-\lambda^{(1)}_{1})}
\frac{1}{b(\lambda-\lambda^{(1)}_{2})}
F(\lambda^{(1)}_{1}) \sum_{ij} \xi_{ji} {\cal F}^{ji}
\EN
where the first term (A.8) comes from the eigenvector part proportional to
$F(\lambda_1^{(1)})$ while the remaining ones originate from the term
$B_{i}(\lambda^{(1)}_{1})B_{j}(\lambda^{(1)}_{2}) {\cal F}^{ji}$. In order to
go on we need to take advantage of the identity
\EQ
\Lambda^{(1)}(\lambda,\{\lambda^{(1)}_{i}\}) {\cal F}^{ml} = 
[T^{(1)}(\lambda,\{\lambda^{(1)}_{i}\})]_{ij}^{lm} {\cal F}^{ji}
\EN

By expanding expressions (A.8-A.10) with the helping of identity (A.11) we have
checked that these three 
terms together produce the third piece of the eigenvalue expression (55).

We now turn our attention to the unwanted terms. They are generated when the 
parameter $\lambda$ of the diagonal operators $B(\lambda)$, $A_{aa}(\lambda)$
and $D(\lambda)$ exchanges with the variables $\{ \lambda_1^{(1)}, \lambda_2^{(1)} \} $ parametrizing the eigenvector 
$\ket{\Phi_{2}(\lambda^{(1)}_{1},\lambda^{(1)}_{2})}$.
Basically, we have three
kinds of such terms and they are proportional to 
\EQ
B_{i}(\lambda)B_{j}(\lambda^{(1)}_{i}),~~
B^{*}_{i}(\lambda)B_{j}(\lambda^{(1)}_{i}),~~
F(\lambda)
\EN

When $\lambda_i^{(1)}=\lambda_2^{(1)}$ we just have two contributions for
the first two unwanted terms (A.12), and they are canceled out by same
procedure presented in section 4
( see equations (51,52) ). By contrast, when 
$\lambda_i^{(1)}=\lambda_1^{(1)}$, several terms appear and the simplifications
are more involved. For example, in the case of the unwanted term 
$B_{i}(\lambda)B_{j}(\lambda^{(1)}_{1}) {\cal{F}}^{ji}$ we have four contributions coming
from 
$B_{i}(\lambda^{(1)}_{1})B_{j}(\lambda^{(1)}_{2})$ 
which are given by
\bear
-[b(\lambda^{(1)}_{2})] \frac{a(\lambda-\lambda^{(1)}_{1})}{b(\lambda-\lambda^{(1)}_{1})}
\frac{1}{b(\lambda-\lambda^{(1)}_{2})} 
\sum_{ij} B_{j}(\lambda)B_{i}(\lambda^{(1)}_{1}) {\cal F}^{ji};
\nonumber \\
+[b(\lambda^{(1)}_{2})]^{L} \frac{1}{b(\lambda-\lambda^{(1)}_{1})}
\frac{1}{b(\lambda^{(1)}_{1}-\lambda^{(1)}_{2})} 
\sum_{ij} B_{j}(\lambda)B_{i}(\lambda^{(1)}_{1}) {\cal F}^{ji};
\nonumber \\
-[a(\lambda^{(1)}_{2})]^{L} \frac{1}{b(\lambda^{(1)}_{1}-\lambda)}
\frac{1}{b(\lambda^{(1)}_{2}-\lambda)} 
\sum_{ijlm}[X^{(1)}(\lambda^{(1)}_{1}-\lambda)]_{lm}^{ij}
 B_{l}(\lambda)B_{m}(\lambda^{(1)}_{1}) {\cal F}^{ji};
\nonumber \\
+[a(\lambda^{(1)}_{2})]^{L} \frac{1}{b(\lambda^{(1)}_{1}-\lambda)}
\frac{1}{b(\lambda^{(1)}_{2}-\lambda^{(1)}_{1})} 
\sum_{ij} B_{i}(\lambda)B_{j}(\lambda^{(1)}_{1}) {\cal F}^{ji}
\ear
and only  one  coming from the term $F(\lambda^{(1)}_{1}) 
\vec{\xi}.\vec{\cal F}$, which is given by
\EQ
[a(\lambda^{(1)}_{2})]^{L} \frac{d_{n}(\lambda^{(1)}_{1}-\lambda)}{e_{n}(\lambda^{(1)}_{1}-\lambda)}
\frac{d_{n}(\lambda^{(1)}_{1}-\lambda^{(1)}_{2})}{e_{n}(\lambda^{(1)}_{1}-\lambda^{(1)}_{2})} 
\sum_{ijlm} \xi^{*}_{ml}B_{l}(\lambda)B_{m}(\lambda^{(1)}_{1}) \xi_{ij} {\cal F}^{ji}
\EN
The simplest way of checking  that the terms (A.13) and (A.14) together 
cancel out is
by using explicitly the Bethe ansatz equations (45) 
for variable $\lambda_2^{(1)}$, namely
\EQ
[b(\lambda^{(1)}_{2})]^{L} {\cal F}^{lm} = 
\frac{b(\lambda^{(1)}_{2}-\lambda^{(1)}_{1})}{b(\lambda^{(1)}_{1}-\lambda^{(1)}_{2})}
\frac{1}{a(\lambda^{(1)}_{2}-\lambda^{(1)}_{1})}
[a(\lambda^{(1)}_{2})]^{L} 
\sum_{ij}[X^{(1)}(\lambda^{(1)}_{1}-\lambda^{(1)}_{2})]_{ij}^{lm} {\cal F}^{ji} 
\EN

Substituting this expression in the first two terms of equation (A.13), all
the unwanted terms become proportional to 
$[a(\lambda^{(1)}_{2})]^L$. Then, we have verified that such
final expression gives a null result with the helping of
some additional identities between the Boltzmann weigths. Analagously, 
a similar 
procedure can be used to deal with the unwanted term 
$B^{*}_{i}(\lambda) B_{j}(\lambda^{(1)}_{1})$. The five contributions for this
unwanted term are collected below
\bear
&& [a(\lambda^{(1)}_{2})]^{L} \frac{1}{b(\lambda-\lambda^{(1)}_{1})}
\frac{d_{n}(\lambda-\lambda^{(1)}_{2})}{e_{n}(\lambda-\lambda^{(1)}_{2})} 
\sum_{aijlm}[X^{(1)}(\lambda-\lambda^{(1)}_{1})]_{ai}^{lm}
 B^{*}_{a}(\lambda)B_{l}(\lambda^{(1)}_{1}) \xi_{mj} {\cal F}^{ji};
\nonumber \\
&& -[a(\lambda^{(1)}_{2})]^{L} \frac{d_{n}(\lambda-\lambda^{(1)}_{1})}
{e_{n}(\lambda-\lambda^{(1)}_{1})}
\frac{1}{b(\lambda^{(1)}_{2}-\lambda^{(1)}_{1})} 
\sum_{aij} \xi_{ai} B^{*}_{a}(\lambda)B_{j}(\lambda^{(1)}_{1}) {\cal F}^{ji};
\nonumber \\
&& -[b(\lambda^{(1)}_{2})]^{L} \left[
\frac{b(\lambda-\lambda^{(1)}_{1})}{e_{n}(\lambda-\lambda^{(1)}_{1})}
\frac{d_{n}(\lambda-\lambda^{(1)}_{2})}{e_{n}(\lambda-\lambda^{(1)}_{2})}
-\frac{d_{n}(\lambda-\lambda^{(1)}_{1})}{e_{n}(\lambda-\lambda^{(1)}_{1})}
\frac{1}{b(\lambda^{(1)}_{1}-\lambda^{(1)}_{2})} \right] 
\sum_{ijl} \xi_{lj} B^{*}_{l}(\lambda)B_{i}(\lambda^{(1)}_{1}) {\cal F}^{ji};
\nonumber \\
&& -[a(\lambda^{(1)}_{2})]^{L} \frac{1}{b(\lambda-\lambda^{(1)}_{1})}
\frac{d_{n}(\lambda^{(1)}_{1}-\lambda^{(1)}_{2})}
{e_{n}(\lambda^{(1)}_{1}-\lambda^{(1)}_{2})} 
\sum_{ijl} \xi_{ij} B^{*}_{l}(\lambda)B_{l}(\lambda^{(1)}_{1}) {\cal F}^{ji}
\ear

Lastly, we have nine contributions to the third term proportional to $F(\lambda)$. The first seven come from
$B_{i}(\lambda^{(1)}_{1})B_{j}(\lambda^{(1)}_{2})$ and the other two  
from $F(\lambda^{(1)}_{1})$. An easy way of verifying that all
these nine terms cancel out is by using explicitly the Bethe ansatz equations (45) both for $\lambda_1^{(1)}$ and $\lambda_2^{(1)}$. The final 
expression can be disentangled only in terms of 
$[a(\lambda_1^{(1)}) a(\lambda_2^{(1)})]^{L} F(\lambda) $, and we have
checked that it gives a null by using $Mathematica^{TM}$.

\centerline{\bf Appendix B : The three-particle state symmetrization }
\setcounter{equation}{0}
\renewcommand{\theequation}{B.\arabic{equation}}

This appendix is concerned with the symmetric properties of the three-particle
state. We begin our discussion with the $\lambda_2^{(1)} 
\leftrightarrow \lambda_3^{(1)}$ permutation. After this permutation the
(vector) three-particle state (46) looks like
\bear
&& \vec{\Phi}_{3}(\lambda^{(1)}_{1},\lambda^{(1)}_{3},\lambda^{(1)}_{2}) =
\vec{B}(\lambda^{(1)}_{1}) \otimes 
\vec{\Phi}_{2}(\lambda^{(1)}_{3},\lambda^{(1)}_{2}) + 
\nonumber \\
&& \left[ \vec{\xi} \otimes  F(\lambda^{(1)}_{1})
 \vec{B}(\lambda^{(1)}_{2}) B(\lambda^{(1)}_{3}) \right]
 \hat{h}_{1}(\lambda^{(1)}_{1},\lambda^{(1)}_{3},\lambda^{(1)}_{2})   
+ \left[ \vec{\xi} \otimes  F(\lambda^{(1)}_{1}) 
\vec{B}(\lambda^{(1)}_{3}) B(\lambda^{(1)}_{2}) \right]   
 \hat{h}_{2}(\lambda^{(1)}_{1},\lambda^{(1)}_{3},\lambda^{(1)}_{2}) 
\nonumber \\
\ear

This vector can be related to  
$\vec{\Phi}_{3}(\lambda^{(1)}_{1},\lambda^{(1)}_{2},\lambda^{(1)}_3)$ by
commuting the fields
$\vec{B}(\lambda^{(1)}_{2})$ and $\vec{B}(\lambda^{(1)}_{3})$ in 
 expression (46). In order to do that, we
have to use the commutation rule (29), which now reads
\bear
\vec{B}(\lambda_2^{(1)}) \otimes \vec{B}(\lambda_3^{(1)}) = 
\frac{1}{a(\lambda_2^{(1)}-\lambda_3^{(1)})}
[ \vec{B}(\lambda_3^{(1)}) \otimes \vec{B}(\lambda_2^{(1)}) ]. 
X^{(1)}(\lambda_2^{(1)}-\lambda_3^{(1)})
\nonumber \\
+\frac{d_{n}(\lambda_2^{(1)}-\lambda_3^{(1)})}{e_{n}(
\lambda_2^{(1)}-\lambda_3^{(1)})}  
\left[ F(\lambda_2^{(1)})B(\lambda_3^{(1)}) - 
\frac{\{ 1-\hat{t} b(\lambda_2^{(1)}-\lambda_3^{(1)}) \}
 }{a(\lambda_2^{(1)}-\lambda_3^{(1)})}F(\lambda_3^{(1)})
B(\lambda_2^{(1)}) \right] \vec{\xi}
\ear

By substituting this relation in equation (46) and by 
comparing the final result with 
the vector (B.1), we find that the symmetric property (50) holds 
for variables $\{ \lambda_2^{(1)}, \lambda_3^{(1)} \} $ provided the functions
$\hat{h}_{1}(\lambda^{(1)}_{1},\lambda^{(1)}_{2},\lambda^{(1)}_{3})$ and 
$\hat{h}_{2}(\lambda^{(1)}_{1},\lambda^{(1)}_{2},\lambda^{(1)}_{3})$  
satisfy
\EQ
\hat{h}_{2}(\lambda^{(1)}_{1},\lambda^{(1)}_{2},\lambda^{(1)}_{3})= \hat{h}_{1}(\lambda^{(1)}_{1},\lambda^{(1)}_{3},\lambda^{(1)}_{2}) 
\frac{X_{23}^{(1)}(\lambda^{(1)}_{2}-\lambda^{(1)}_{3})}{
a(\lambda^{(1)}_{2} - \lambda^{(1)}_{3}) }
\EN
\EQ
\hat{h}_{1}(\lambda^{(1)}_{1},\lambda^{(1)}_{2},\lambda^{(1)}_{3})= \hat{h}_{2}(\lambda^{(1)}_{1},\lambda^{(1)}_{3},\lambda^{(1)}_{2}) 
\frac{X_{23}^{(1)}(\lambda^{(1)}_{2}-\lambda^{(1)}_{3})}{
a(\lambda^{(1)}_{2} - \lambda^{(1)}_{3}) }
\EN

In fact, the two expressions above are equivalent due to the unitarity 
property of the auxiliary matrix $X^{(1)}(\lambda)$, namely
\EQ
X^{(1)}(u) X^{(1)}(-u) = a(u)a(-u)  I
\EN

Therefore, in order to determine these functions we still need to find an extra 
constraint. This comes out if we perform the permutation between 
variables $\lambda_1^{(1)}$ and $\lambda_2^{(1)}$. The procedure is similar 
to the one described above, but now some additional commutation rules are
needed. For example, besides commuting the creation
fields $\vec{B}(\lambda_1^{(1)})$
and $\vec{B}(\lambda_2^{(1)})$, we have to commute $\vec{B}(\lambda_1^{(1)})$
and  $F(\lambda_2^{(1)})$ and also
$F(\lambda_1^{(1)})$ and $\vec{B}(\lambda_2^{(1)})$. This latter step is 
worked out with the helping of commutations rules (31) and (32), respectively.
We then are able to determine the function 
$\hat{h}_{1}(\lambda^{(1)}_{1},\lambda^{(1)}_{2},\lambda^{(1)}_{3})$ by
imposing the necessary restriction that the terms proportional to
$ \vec{\xi} \otimes  F(\lambda^{(1)}_{1})\vec{B}(\lambda^{(1)}_{3}) B(\lambda^{(1)}_{2})$  
need to be canceled out. We find that this condition is verified provided
\EQ
\hat{h}_{1}(\lambda^{(1)}_{1},\lambda^{(1)}_{2},\lambda^{(1)}_{3}) = 
- \frac{d_n(\lambda^{(1)}_{1}-\lambda^{(1)}_{2})}{e_n(\lambda^{(1)}_{1}-\lambda^{(1)}_{2})}  
 \frac{a(\lambda^{(1)}_{3}-\lambda^{(1)}_{2})}{b(\lambda^{(1)}_{3}-\lambda^{(1)}_{2})} I
\EN
and by substituting (B.6) in (B.3) we are able to fix 
$\hat{h}_{2}(\lambda^{(1)}_{1},\lambda^{(1)}_{2},\lambda^{(1)}_{3})$ ( see equation (48) ).

Besides that, some other non-trivial checks need to be performed in 
order to verify the consistency of the procedure mentioned above. In 
the process of using the three commutation rules mentioned above, certain
extra
identities emerge, and they are given by
\begin{eqnarray}
& [ \vec{\xi} \otimes \vec{B}(\lambda^{(1)}_{2}) 
F(\lambda^{(1)}_{1}) B(\lambda^{(1)}_{3}) ] b(\lambda^{(1)}_{1}-\lambda^{(1)}_{2})
\left[
-\frac{1}{b(\lambda^{(1)}_{3}-\lambda^{(1)}_{2})} 
\frac{d_{n}(\lambda^{(1)}_{1}-\lambda^{(1)}_{2})}
{e_{n}(\lambda^{(1)}_{1}-\lambda^{(1)}_{2})} + 
\hat{h}_{2}(\lambda^{(1)}_{1},\lambda^{(1)}_{2},\lambda^{(1)}_{3}) \right] =
\nonumber \\
&\frac{d_{n}(\lambda^{(1)}_{2}-\lambda^{(1)}_{3})}
{e_{n}(\lambda^{(1)}_{2}-\lambda^{(1)}_{3})}
[ \vec{B}(\lambda^{(1)}_{2}) F(\lambda^{(1)}_{1})
 B(\lambda^{(1)}_{3}) \otimes \vec{\xi} ]
-\frac{d_{n}(\lambda^{(1)}_{1}-\lambda^{(1)}_{3})}
{e_{n}(\lambda^{(1)}_{1}-\lambda^{(1)}_{3})}
[ \vec{B}(\lambda^{(1)}_{2}) F(\lambda^{(1)}_{1})
 B(\lambda^{(1)}_{3}) \otimes \vec{\xi} ] 
X_{12}^{(1)}(\lambda^{(1)}_{1}-\lambda^{(1)}_{2}) 
\nonumber \\
\end{eqnarray}
and
\bear
& [ \vec{\xi} \otimes F(\lambda^{(1)}_{2})
 \vec{B}(\lambda^{(1)}_{1}) B(\lambda^{(1)}_{3}) ] 
\left[ 
\frac{[1-\hat{t} b(\lambda_1^{(1)}-\lambda_2^{(1)})]}{b(\lambda^{(1)}_{3}-\lambda^{(1)}_{1})} 
\frac{d_{n}(\lambda^{(1)}_{1}-\lambda^{(1)}_{2})}
{e_{n}(\lambda^{(1)}_{1}-\lambda^{(1)}_{2})}
-\frac{1}{b(\lambda^{(1)}_{3}-\lambda^{(1)}_{2})} 
\frac{d_{n}(\lambda^{(1)}_{1}-\lambda^{(1)}_{2})}
{e_{n}(\lambda^{(1)}_{1}-\lambda^{(1)}_{2})} \right]
\nonumber \\
& -b(\lambda^{(1)}_{1}-\lambda^{(1)}_{2}) 
\frac{d_{n}(\lambda^{(1)}_{2}-\lambda^{(1)}_{3})}
{e_{n}(\lambda^{(1)}_{2}-\lambda^{(1)}_{3})}
[ F(\lambda^{(1)}_{2}) \vec{B}(\lambda^{(1)}_{1})
 B(\lambda^{(1)}_{3})  \otimes  \vec{\xi} ] =
\nonumber \\
& [ \vec{\xi} \otimes F(\lambda^{(1)}_{2}) 
\vec{B}(\lambda^{(1)}_{1}) B(\lambda^{(1)}_{3}) ] 
\left[ 
-\hat{h}_{2}(\lambda^{(1)}_{1},\lambda^{(1)}_{2},\lambda^{(1)}_{3}) +
X_{12}^{(1)}(\lambda^{(1)}_{1}-\lambda^{(1)}_{2})
 \hat{h}_{2}(\lambda^{(1)}_{2},\lambda^{(1)}_{1},\lambda^{(1)}_{3}) \right]
\ear 

In order to show that these relations are satisfied, we end up proving
remarkable properties between 
$\vec{\xi}$  and the auxiliary matrix
$X^{(1)}(u)$. They are listed below as
\EQ
[\vec{\xi}_{12} \vec{B}_{3}(y)] X_{12}^{(1)}(u) = 
\frac{d_n(u)}{e_n(u)} \frac{e_n(-u)}{d_n(-u)}[1-\hat{t}b(u)] 
[\vec{\xi}_{12} \vec{B}_{3}(y)] 
\EN
\EQ
[\vec{B}_{1}(y) \vec{\xi}_{23}] X_{23}^{(1)}(u) =
\frac{d_n(u)}{e_n(u)} \frac{e_n(-u)}{d_n(-u)}[1-\hat{t}b(u)] 
 [\vec{B}_{1}(y) \vec{\xi}_{23}]
\EN
\EQ
[\vec{B}_{1}(y) \vec{\xi}_{23}] X_{12}^{(1)}(u) =
[\vec{B}_{1}(y) \vec{\xi}_{23}] + u [\vec{B}_{1}(y) 
\vec{\xi}_{23} P^{(1)g}_{12}] 
-\frac{u}{u-\Delta^{(1)}} [\vec{\xi}_{12} \vec{B}_{3}(y)]
\EN
\EQ
[\vec{\xi}_{12} \vec{B}_{3}(y)]  X_{23}^{(1)}(u) =
[\vec{\xi}_{12} \vec{B}_{3}(y)] + u [\vec{\xi}_{12} 
\vec{B}_{3}(y) P^{(1)g}_{23}]
-\frac{u}{u-\Delta^{(1)}} [\vec{B}_{1}(y) \vec{\xi}_{23}]
\EN
\EQ 
[\vec{\xi}_{12} \vec{B}_{3}(y) P^{(1)g}_{23}]  X_{12}^{(1)}(u) = 
[\vec{\xi}_{12} \vec{B}_{3}(y) P^{(1)g}_{23}] + u [\vec{B}_{1}(y) \vec{\xi}_{23}] 
-\frac{u}{u-\Delta^{(1)}} \hat{t} [\vec{\xi}_{12} \vec{B}_{3}(y)]
\EN
where the lower indices indicate the position where vector $\vec{\xi}$,
vector $\vec{B}(y)$ and matrix $X^{(1)}(u)$ acts in a non-trivial way. 
$P^{(1)g}$ means the permutator entering in the construction of $X^{(1)}(\lambda)$.
Besides that, we also note the following 
useful identity $\vec{\xi}_{12} \vec{B}_{3}(y) P^{(1)g}_{23}=
\vec{B}_{1}(y) \vec{\xi}_{23} P^{(1)g}_{12}$.

\centerline{\bf Appendix C : The graded quantum inverse approach }
\setcounter{equation}{0}
\renewcommand{\theequation}{C.\arabic{equation}}

The purpose of this appendix is to  discuss the main modifications occurring
in the commutation rules, eigenvalues and Bethe ansatz equations for the
models  
 $Osp(2n-1|2)$, $Osp(2|2n-2)$, $Osp(2n-2|2)$  and 
$Osp(1|2n)$ when  we formulate their solution in terms of the graded inverse
scattering framework \cite{KS}. In this formalism, the vertex operator 
${\cal L}_{{\cal A} i }(\lambda) $ entering in the monodromy matrix (12) 
satisfies the graded Yang-Baxter equation and is given by acting the graded
permutation operator $P^{g}$ on the $R$-matrix (4), namely
\EQ
{\cal L}_{ab}^{cd}(\lambda) = (-1)^{p(a)p(b)} R^{cd}_{ab}(\lambda)
\EN

In order to accomplish this change, the integrability condition (14) now reads
\EQ
R(\lambda , \mu) {\cal T}(\lambda)  \stackrel{s}{\otimes} {\cal T}(\mu) =
{\cal T}(\mu)  \stackrel{s}{\otimes} {\cal T}(\lambda)  R(\lambda ,\mu).
\EN	
where the symbol
$ {\stackrel{s}{\otimes}} $ stands for the supertensor product 
$
\left( A \stackrel{s}{\otimes} B \right)_{ab}^{cd} = 
(-1)^{p(b)[p(a)+p(c)]} A_{ac} B_{bd}
$. Moreover, the transfer matrix $T(\lambda)$ is written in terms of a
supertrace of the monodromy matrix $\cal{T} (\lambda)$ as
\EQ
T(\lambda) = Str {\cal T}(\lambda) = 
\sum_{a} (-1)^{p(a)} {\cal T}_{aa}(\lambda) 
\EN

The modifications $(C.1-C.3)$ are responsible by the appearance of extra
signs in some terms of the commutation rules, in the ``diagonal'' conditions
(19) and in the eigenvalue problem (18). Furthermore, the way that 
such signs enter
on these relations depend much on the grading sequence we have chosen. We
recall that we have taken the $fb \cdots bf$ grading for the first three models
and $b \cdots bfb \cdots b$ for the $Osp(1|2n)$ vertex model. Therefore, 
we have
to present the modifications for these two classes of grading separately. We
begin by listing the results for the 
$Osp(2n-1|2)$, $Osp(2|2n-2)$ and $Osp(2n-2|2)$ models. Concerning commutation
rules, the extra signs appear only in relations (22,23,24,29) 
and for the following
right hand side pairs of terms
\EQ
B_{i} B_{j} \rightarrow - B_{i}B_{j}, ~~
B_{i} B \rightarrow - B_{i} B, ~~
B_{i} D \rightarrow - B_{i} D, ~~
B^{*}_{i} B \rightarrow - B^{*}_{i} B.
\EN

All the other fundamental commutation relations  remain
unchanged. Furthermore, the action of the diagonal operators $B(\lambda)$,
$A_{ab}(\lambda)$ and $D(\lambda)$ on the reference state (15) becomes
\EQ
B(\lambda)\ket{0} = [-a(\lambda)]^{L}\ket{0},~~ D(\lambda)\ket{0} = 
[-e_{n}(\lambda)]^{L}\ket{0},~~ A_{aa}(\lambda)\ket{0} = [b(\lambda)]^{L}\ket{0} , a=1, \dots , q-2.
\EN
while the eigenvalue problem (18) becomes
\EQ
[-B(\lambda)+\sum_{a=1}^{q-2}A_{aa}(\lambda)-D(\lambda)] \ket{\Phi} = 
\Lambda(\lambda) \ket{\Phi}
\EN

Taking into account these relations one can easily verify that the phase factor
$(-1)^{L-m_1-1}$ present in the Bethe ansatz equations of these models is
canceled out. For sake of completeness we also present the final results for
the eigenvalues in the graded formalism. 
For $Osp(2n-1|2)$ ($n \geq 2)$ we have
\bear
&& \Lambda^{Osp(2n-1|2)} \left(\lambda;\{\lambda^{(1)}_j \}, 
\cdots,\{\lambda^{(n)}_j\} \right) = 
\nonumber \\
&& -[-a(\lambda)]^L \prod_{j=1}^{m_1} \frac{\lambda-\lambda^{(1)}_{j}+\frac{1}{2}}
{\lambda-\lambda^{(1)}_{j}-\frac{1}{2}} -
[-e_n(\lambda)]^L \prod_{j=1}^{m_1} \frac{\lambda-\lambda^{(1)}_{j}+n-3 }
{\lambda-\lambda^{(1)}_{j}+n-2} +
[b(\lambda)]^L \sum_{l=1}^{2n-1} G_{l}(\lambda,\{ \lambda_j^{(\beta)} \}  )
\nonumber \\
\ear
and for  $Osp(2|2n-2)$, $(n \geq 2)$ we find
\bear
&& \Lambda^{Osp(2|2n-2)} \left(\lambda;\{\lambda^{(1)}_j \}, 
\cdots,\{\lambda^{(n)}_j\} \right) = 
\nonumber \\
&& -[-a(\lambda)]^L \prod_{j=1}^{m_1} \frac{\lambda-\lambda^{(1)}_{j}+\frac{1}{2}}
{\lambda-\lambda^{(1)}_{j}-\frac{1}{2}} -
[-e_n(\lambda)]^L \prod_{j=1}^{m_1} \frac{\lambda-\lambda^{(1)}_{j}+\frac{2n-3}{2}}
{\lambda-\lambda^{(1)}_{j}+\frac{2n-1}{2}} +
[b(\lambda)]^L \sum_{l=1}^{2n-2} G_{l}(\lambda,\{\lambda_j^{(\beta)} \})
\nonumber \\
\ear
while for $Osp(2n-2|2)$ ($n\geq 3)$ we have
\bear
&& \Lambda^{Osp(2n-2|2)} \left( \lambda;\{\lambda^{(1)}_j \}, 
\cdots,\{\lambda^{(n)}_j\} \right) = \nonumber \\
&& -[-a(\lambda)]^L \prod_{j=1}^{m_1} \frac{\lambda-\lambda^{(1)}_j+\frac{1}{2}}
{\lambda- \lambda^{(1)}_j-\frac{1}{2}} -
[-e_n(\lambda)]^L \prod_{j=1}^{m_1} -\frac{\lambda-\lambda^{(1)}_j+\frac{2n-7}{2}}
{\lambda- \lambda^{(1)}_j+\frac{2n-5}{2}} 
 + [b(\lambda)]^L \sum_{l=1}^{2n-2} G_{l}(\lambda,\{ \lambda_j^{(\beta)} \})
\nonumber \\
\ear
where functions $G_{l}(\lambda,\{\lambda_j^{(\beta)})$ remain unchanged and
are given by equations (87), (89) and (91), respectively.

The analysis for  $Osp(1|2n)$ model is  a bit simpler because we just
have one ``central'' fermionic species in the grading sequence. In this case,
we have only extra signs for the terms $B_i A_{jl}$ and $[B^{*}]_i A_{jl} $
in  relations (22) and (24), respectively. The change in the eigenvalue (92)
now appears on the $n$-th term, and the final result is 
\bear
&& \Lambda^{Osp(1|2n)} \left(\lambda;\{\lambda^{(1)}_j \}, 
\cdots,\{\lambda^{(n)}_j\} \right) = 
\nonumber \\
&& [a(\lambda)]^L \prod_{j=1}^{m_1} \frac{\lambda-\lambda^{(1)}_{j}-\frac{1}{2}}
{\lambda-\lambda^{(1)}_{j}+\frac{1}{2}} +
[e_n(\lambda)]^L \prod_{j=1}^{m_1} \frac{\lambda-\lambda^{(1)}_{j}+n+1}
{\lambda-\lambda^{(1)}_{j}+n} +
\nonumber \\
&& [b(\lambda)]^L \sum_{l=1}^{n-1} G_{l}(\lambda,\{ \lambda_j^{(\beta)} \})+
[b(\lambda)]^L \sum_{l=n+1}^{2n-1} G_{l}(\lambda,\{ \lambda_j^{(\beta)} \})
-[b(\lambda)]^L  {\tilde{G}}_{n}(\lambda,\{ \lambda_j^{(\beta)} \})
\ear
where functions $G_l(\lambda,\{\lambda_j^{(\beta)}\}$ for $l=1,\cdots,n-1,n+1,
\cdots, 2n-1$ are still given by (93), but ${\tilde{G}}_n(\lambda,\{ \lambda_j^{(\beta)} \})$ is
\EQ
 {\tilde{G}}_{n}(\lambda,\{ \lambda_j^{(\beta)}\}) = 
 \prod_{k=1}^{m_{(n)}} \frac{(\lambda-\frac{\lambda^{(n)}_{k}}{i}+\frac{(n-1)}{2})}
{(\lambda-\frac{\lambda^{(n)}_{k}}{i}+\frac{(n+1)}{2})}
\frac{(\lambda-\frac{\lambda^{(n)}_{k}}{i}+\frac{(n+2)}{2})} 
{(\lambda-\frac{\lambda^{(n)}_{k}}{i}+\frac{n}{2})}
\EN

Analogously, the Bethe ansatz equations only 
modify for the last root $\{\lambda_j^{(n)} \}$. Instead of 
equation (104) we now have
\EQ
\prod_{k=1}^{m_{n-1}} \frac{\lambda^{(n)}_{j}-\lambda^{(n-1)}_{k}+1/2}{\lambda^{(n)}_{j}-\lambda^{(n-1)}_{k}-1/2} 
\prod_{k=1}^{m_{n}} \frac{(\lambda^{(n)}_{j}-\lambda^{(n)}_{k}+1/2)(\lambda^{(n)}_{j}-\lambda^{(n)}_{k}-1)}
{(\lambda^{(n)}_{j[}-\lambda^{(n)}_{k}-1/2)(\lambda^{(n)}_{j}-\lambda^{(n)}_{k}+1)} = 1
\EN
%


\newpage
     
\Large
Tables 
\normalsize
\vspace{1.5cm}

\underline{Table 1}: The number of possible states per link $q$ and the parameters $\hat{t}$ and $\cal{K}$ for the models 
$B_{n}$, $C_{n}$, $D_{n}$, $Osp(2n-1|2)$, $Osp(2|2n-2)$, $Osp(2n-2|2)$ and $Osp(1|2n)$.

\btb[h]
\bc
\bt{|c|c|c|c|c|c|c|c|} \hline
     &$B_{n}$  &$C_{n}$  &$D_{n}$  &$Osp(2n-1|2)$  &$Osp(2|2n-2)$  &$Osp(2n-2|2)$  &$Osp(1|2n)$   \\ \hline\hline
$q$   &$2n+1$  &$2n$  &$2n$  &$2n+1$  &$2n$  &$2n$  &$2n+1$          \\ \hline
${\cal K}$  &$2n+1$  &$-2n$  &$2n$  &$2n-3$  &$4-2n$  &$2n-4$  &$1-2n$         \\ \hline
$\hat{t}$   &1  &$-1$  &1  &1  &$-1$  &1  &$-1$          \\ \hline
\et
\ec
\etb
\vspace{1.5cm}

\underline{Table 2}: The main five Boltzmann weights for the vertex models 
$B_{n}$, $C_{n}$, $D_{n}$, $Osp(2n-1|2)$, $Osp(2|2n-2)$,$Osp(2n-2|2)$ and $Osp(1|2n)$.

\btb[h]
\bc
\bt{|c|c|c|c|c|c|c|c|} \hline
     &$B_{n}$  &$C_{n}$  &$D_{n}$  &$Osp(2n-1|2)$  &$Osp(2|2n-2)$  &$Osp(2n-2|2)$  &$Osp(1|2n)$   \\ \hline\hline

$a(\lambda)$  &$\lambda+1$  &$\lambda+1$  &$\lambda+1$  &$1-\lambda$  &$1-\lambda$  &$1-\lambda$  &$\lambda+1$          \\ \hline
$b(\lambda)$  &$\lambda$  &$\lambda$  &$\lambda$  &$\lambda$  &$\lambda$  &$\lambda$  &$\lambda$         \\ \hline
$c_{n}(\lambda)$  &$\frac{n-1/2}{\lambda+n-1/2}$  &$\frac{2\lambda+n+1}{\lambda+n+1}$  
&$\frac{n-1}{\lambda+n-1}$  &$\frac{2\lambda+n-5/2}{\lambda+n-5/2}$  &$\frac{n-1}{\lambda+n-1}$  
&$\frac{2\lambda+n-3}{\lambda+n-3}$  &$\frac{2\lambda+n+1/2}{\lambda+n+1/2}$          \\ \hline
$d_{n}(\lambda)$  &$\frac{-\lambda}{\lambda+n-1/2}$  &$\frac{\lambda}{\lambda+n+1}$  
&$-\frac{\lambda}{\lambda+n-1}$  &$\frac{-\lambda}{\lambda+n-5/2}$  &$\frac{\lambda}{\lambda+n-1}$  
&$\frac{-\lambda}{\lambda+n-3}$  &$-\frac{\lambda}{\lambda+n+1/2}$        \\ \hline
$e_{n}(\lambda)$  &$\frac{\lambda(\lambda+n-3/2)}{\lambda+n-1/2}$  &$\frac{\lambda(\lambda+n)}{\lambda+n+1}$  
&$\frac{\lambda(\lambda+n-2)}{\lambda+n-1}$  &$\frac{-\lambda(\lambda+n-3/2)}{\lambda+n-5/2}$  
&$-\frac{\lambda(\lambda+n)}{\lambda+n-1}$  &$\frac{-\lambda(\lambda+n-2)}{\lambda+n-3}$   
&$\frac{\lambda(\lambda+n-1/2)}{\lambda+n+1/2}$        \\ \hline
\et
\ec
\etb
\newpage
\vspace{1.5cm}
\underline{Table 3}: The main five Boltzmann weights 
of `nested' matrix $X^{(k)}$. They are the same for the pairs  
$\{ B_{n}, Osp(2n-1|2) \}$, $\{ C_{n}, Osp(2|2n-2) \}$ and  
$\{D_{n}, Osp(2n-2|2) \}$. The corresponding crossing parameter $\Delta^{(k)}$ 
is also listed. 

\btb[h]
\bc
\bt{|c|c|c|c|c|c|} \hline
$X^{(k)}(\lambda)$    &$\{ B_{n},Osp(2n-1|2) \} $  &$\{ C_{n}, Osp(2|2n-2) \}$  &$\{ D_{n}, Osp(2n-2|2) \}$  &$Osp(1|2n)$     \\ \hline\hline

$a(\lambda)$  &$\lambda+1$  &$\lambda+1$  &$\lambda+1$    &$\lambda+1$          \\ \hline
$b(\lambda)$  &$\lambda$  &$\lambda$  &$\lambda$    &$\lambda$        \\ \hline
$c_{n-k}(\lambda)$  &$\frac{n-k-1/2}{\lambda+n-k-1/2}$  &$\frac{2\lambda+n-k+1}{\lambda+n-k+1}$  
&$\frac{n-k-1}{\lambda+n-k-1}$  &$\frac{2\lambda+n-k+1/2}{\lambda+n-k+1/2}$        \\ \hline
$d_{n-k}(\lambda)$  &$\frac{-\lambda}{\lambda+n-k-1/2}$  &$\frac{\lambda}{\lambda+n-k+1}$  
&$\frac{-\lambda}{\lambda+n-k-1}$   &$-\frac{\lambda}{\lambda+n-k+1/2}$        \\ \hline
$e_{n-k}(\lambda)$  &$\frac{\lambda(\lambda+n-k-3/2)}{\lambda+n-k-1/2}$  &$\frac{\lambda(\lambda+n-k)}{\lambda+n-k+1}$  
&$\frac{\lambda(\lambda+n-k-2)}{\lambda+n-k-1}$  &$\frac{\lambda(\lambda+n-k-1/2)}{\lambda+n-k+1/2}$        \\ \hline
$\Delta^{(k)}$  &$-n+k+1/2$  &$-n+k-1$  &$-n+k+1$  &$-n+k-1/2 $           \\ \hline
\et
\ec
\etb
\underline{Table 4}: The shifts $\delta^{(\beta)}$ performed on the
Bethe Ansatz variables $\lambda_j^{(\beta)} $.

\btb[h]
\bc
\bt{|c|c|c|c|} \hline
     &$B_{n}$ and $Osp(1|2n)$  &$C_{n}$  &$D_{n}$     \\ \hline\hline
$\delta^{(\beta)}$   &$\beta/2,~~ \beta=1,\cdots, n$  
&$ \cases{ \beta/2, & $\beta=1,\cdots, n-1$ \cr
(n+1)/2, & $\beta=n$ \cr }$  
& $\cases{ \beta/2,& $\beta=1,\cdots,n-2$ \cr
(n-1)/2, & $\beta= \pm $ \cr}$  
\\ \hline
\et
\ec
\etb
\vspace{0.1cm}

\btb[h]
\bc
\bt{|c|c|c|c|} \hline
     &$Osp(2n-1|2)$  &$Osp(2|2n-2)$  &$Osp(2n-2|2)$     \\ \hline\hline
$\delta^{(\beta)}$    
& $(\beta-2)/2,~~ \beta=1,\cdots,n $  
&$ \cases{(\beta-2)/2, & $\beta=1,\cdots,n-1$ \cr
(n-1)/2, & $\beta=n $ \cr }$  
& $\cases{(\beta-2)/2,& $\beta=1, \cdots, n-2$ \cr
(n-3)/2,&  $\beta= \pm $ \cr }$  \\ \hline
\et
\ec
\etb

\newpage
\centerline{\bf Figure }
\vspace{0.2cm} 
Fig. 1. The  Dynkin diagrams of the Lie algebras 
$B_{n}$, $C_{n}$ and $D_{n}$  and of the superalgebra
$Osp(n|2m)$ (e.g see ref. \cite{CO} ). The  symbols
$\bigcirc $, $\otimes $ and
$\large{\bullet} $  stand for the simple roots of the algebras $Sl(2)$, $Sl(1|1)$ 
and $Osp(1|2)$, respectively. \\

\end{document}